\documentclass[fleqn,11pt]{elsarticle}

\usepackage[utf8]{inputenc}
\usepackage[T1]{fontenc}
\usepackage{mathtools}
\usepackage[letterpaper, margin=0.75in]{geometry}
\usepackage{xcolor}
\usepackage{enumitem}
\usepackage[scr=rsfs]{mathalpha}
\usepackage{mathtools}
\usepackage{amsmath,amsthm,amssymb,amsfonts}
\usepackage{color}
\usepackage{color,soul}
\usepackage{graphicx}
\usepackage{lipsum}
\usepackage{float}
\usepackage{verbatim}
\usepackage{todonotes}
\usepackage{physics}
\usepackage{chemformula}
\usepackage{algorithm,algpseudocode}
\usepackage{siunitx}
\usepackage{booktabs}
\usepackage{tabularx}
\usepackage{svg}
\svgsetup{inkscapelatex=false}

\DeclareSIUnit\Molar{\textsc{M}}

\algnewcommand{\Inputs}[1]{%
  \State \textbf{Inputs:}
  \Statex \hspace*{\algorithmicindent}\parbox[t]{.8\linewidth}{\raggedright #1}
}
\algnewcommand{\Initialize}[1]{%
  \State \textbf{Initialize:}
  \Statex \hspace*{\algorithmicindent}\parbox[t]{.8\linewidth}{\raggedright #1}
}

\makeatletter
\let\ps@pprintTitle\ps@plain
\makeatother
\begin{document}

\begin{frontmatter}

\title{Boundary-contact repolarization provides a mechanism for directional guidance of neural crest cell collectives}


\author[1]{Aditya Shankar Paspunurwar}
\author[1,2,3]{Hector Gomez\corref{cor1}}
\ead{hectorgomez@purdue.edu}
\cortext[cor1]{Corresponding author}

\address[1]{School of Mechanical Engineering, Purdue University, 585 Purdue Mall, West Lafayette, IN 47907, USA}
\address[2]{Weldon School of Biomedical Engineering, Purdue University, 206 S.\ Martin Jischke Drive, West Lafayette, IN 47907, USA}
\address[3]{Purdue Institute for Cancer Research, Purdue University, 201 S.\ University Street, West Lafayette, IN 47907, USA}

\begin{abstract}

Neural crest cells (NCCs) provide a well-established system for investigating collective migration during embryonic development and exhibit migratory behaviors with parallels to cancer invasion.
Previous studies establish roles for co-attraction, contact inhibition of locomotion, polarity persistence, and tissue confinement in guiding NCC migration.
However, it remains unclear whether NCC migration in confined environments occurs through spatial restriction alone or also through a repolarization response induced by contact with boundaries of confinement.
By developing a computational model that couples self-generated chemical transport with contact-mediated repolarization and mechanical interactions, we show that cell-boundary contact can reorient cell polarity and sustain directional migration under confinement, whereas no persistent collective direction emerges without confinement.
Moreover, suppressing contact-mediated repolarization from non-permissive confinement boundaries while preserving the mechanical interactions prevents sustained directional migration, suggesting that spatial restriction alone is insufficient in the model.
%
%
Together, these findings establish an experimentally testable mechanism through which self-generated chemical signals and contact with confining tissues jointly regulate collective guidance of NCCs.
These insights provide a basis for investigating developmental migration defects and potential strategies to regulate collective tissue invasion.


\end{abstract}


\end{frontmatter}



\section{Introduction}

Neural crest cells (NCCs) undergo an epithelial-to-mesenchymal transition (EMT) and migrate over long distances before differentiating into diverse cell types \cite{thiery2009epithelial,mayor1995induction,labonne1998induction,leathers2022time}.
Related EMT programs and collective behaviors also contribute to cancer metastasis during invasion and dissemination \cite{szabo2016modelling,theveneau2013collective}.
Directional migration of NCCs involves mechanisms that preserve collective cohesion, reorient motion after contact with neighboring cells, and constrain migration within surrounding tissues \cite{carmona2008contact,carmona2011complement,szabo2016vivo}.
\textit{Xenopus} NCCs provide a well-characterized system for studying these mechanisms \cite{szabo_mechanisms_2018,sauka2008evolution}.
They secrete the complement fragment C3a, which activates C3aR on nearby cells and promotes co-attraction (CoA) of NCCs \cite{carmona2011complement}.
Contact between neighboring cells also triggers contact inhibition of locomotion (CIL), during which protrusions at the contact interface retract and new protrusions form away from contact \cite{abercrombie1953observations,abercrombie1954observations,carmona2008contact}.
In \textit{Xenopus} NCCs, this response is governed by N-cadherin dynamics, noncanonical Wnt signaling, and redistribution of Rac1 and RhoA activity along the cell membrane \cite{carmona2008contact,ridley2015rho,roycroft2016molecular,raftopoulou2004cell,zegers2014rho}.
Therefore, CoA and cell-cell CIL jointly regulate collective cohesion, spacing and local rearrangement within the migrating population.
\\

\noindent Experiments in which N-cadherin signaling is inhibited and external chemotactic guidance is absent show that NCCs form protrusions without maintaining a preferred orientation \cite{theveneau2010collective}.
Additionally, self-generated C3a attracts cells toward one another but does not define a persistent migration direction in the surrounding tissue \cite{carmona2011complement}.
%
%
However, NCCs are known to migrate through streams bordered by nonpermissive tissues, and versican-enriched boundaries confine these streams and enhance directional persistence \cite{szabo2016vivo}. 
These findings establish an important role for confining tissues in collective guidance of NCCs, but they do not directly distinguish passive spatial restriction from a polarity response triggered by local cell-boundary contact. 
The relative contributions of boundary contact, CoA, cell-cell CIL, and spontaneous motility to the selection and maintenance of a collective migration direction remain difficult to separate, particularly without an external chemotactic gradient. 
\\

Computational models explain the emergence and persistence of collective NCC migration. 
A discrete element model showed that CIL and CoA can jointly produce persistent collective motion under confinement without an externally imposed chemoattractant gradient \cite{woods2014directional}.
A cellular Potts model demonstrated the role of rear boundary in selection of initial migration direction, and reinforced recent cell displacements through a polarity update rule promoting persistence \cite{szabo2016vivo}. 
Models resolving membrane Rac1 and RhoA dynamics provide a biochemical explanation for polarity persistence, with CoA, cell-cell CIL, and CIL-like boundary response suppressing competing protrusions \cite{merchant2018rho,merchant2020rho}.
Related models show that chemical regulation of CIL and interactions between distinct cell groups can generate collective chemotaxis and chase-and-run migration \cite{camley2016emergent,colombi2020modelling}.
\\

However, the effects of changing extracellular environment on chemical guidance remain less resolved in these approaches.
Self-generated chemicals are often represented by analytical concentration profiles \cite{woods2014directional,merchant2018rho,camley2016collective}, while other formulations evolve chemoattractant fields over the computational domain without explicitly excluding cell interiors \cite{szabo2016vivo,colombi2020modelling}.
Under confinement, cell deformation and rearrangement can restrict extracellular transport pathways and alter local C3a accumulation.
The resulting gradients may reinforce or oppose contact-induced repolarization, rather than solely promote cohesion.
Resolving this coupling is therefore important for determining when self-generated chemotaxis and boundary contact sustain a common migration direction and when their competition leads to cellular trapping or migration reversal.
\\

\noindent
Here, we develop a mechanistic model of deformable NCCs that couples C3a secretion and transport, contact-mediated repolarization, stochastic motility, and mechanical interactions among cells and confining boundaries.
The model builds on the dynamic cluster field (DCF) framework \cite{paspunurwar2024dynamic} and represents each cell using a phase field \cite{levine2006directional,emmerich2003diffuse,gomez2017computational,chen1998fiber,palmieri2015multiple,moure2021phase}, thereby resolving cell shape and deformation during migration and contact.
The DCF formulation enables efficient simulation of multicellular populations while preserving individual-cell phase-field dynamics.
We restrict chemical transport to the extracellular space, such that cell motion and deformation continuously alter the available transport environment and the resulting C3a distribution.
A shared contact activity field captures responses at cell-cell and cell-boundary interfaces and generates the corresponding polarity response contributions, while adhesion and repulsion are represented separately.
This formulation allows us to examine how time-evolving extracellular geometry, self-generated chemotaxis, contact-mediated repolarization, and tissue confinement combine to regulate collective behavior of NCCs.
\\

\noindent
We first constrain key mechanical and CIL parameters using experimentally measured pairwise collision dynamics \cite{scarpa2013novel}.
We then use conditions corresponding to C3aR and N-cadherin morpholino (MO) treatments to determine how CoA and cell-cell CIL regulate cohesion and local organization on an unconfined planar substrate.
Next, we study NCC migration in confined corridors, while varying collective size and comparing tight confinement with a widened corridor.
%
%
The simulations reveal that under confinement, boundary-contact repolarization overcomes an initially rear-directed chemotactic bias and provides the outlet-directed cue, while stochastic polarity can reinforce or oppose this response.
Collective size modulates the robustness of this balance, and a widened corridor permits either persistent migration through continued association with a lateral boundary or loss of direction after complete detachment.
Together, these results identify how evolving C3a transport and dynamic contact with boundaries separate collective cohesion from directional guidance.
They also suggest that confining tissue boundaries can actively regulate collective guidance, with implications in embryonic morphogenesis and confined cancer invasion.
\\

\section{Results}

\subsection{\textbf{Model overview}}

\begin{figure}[h!]
    \centering
    \includegraphics[width=\linewidth]{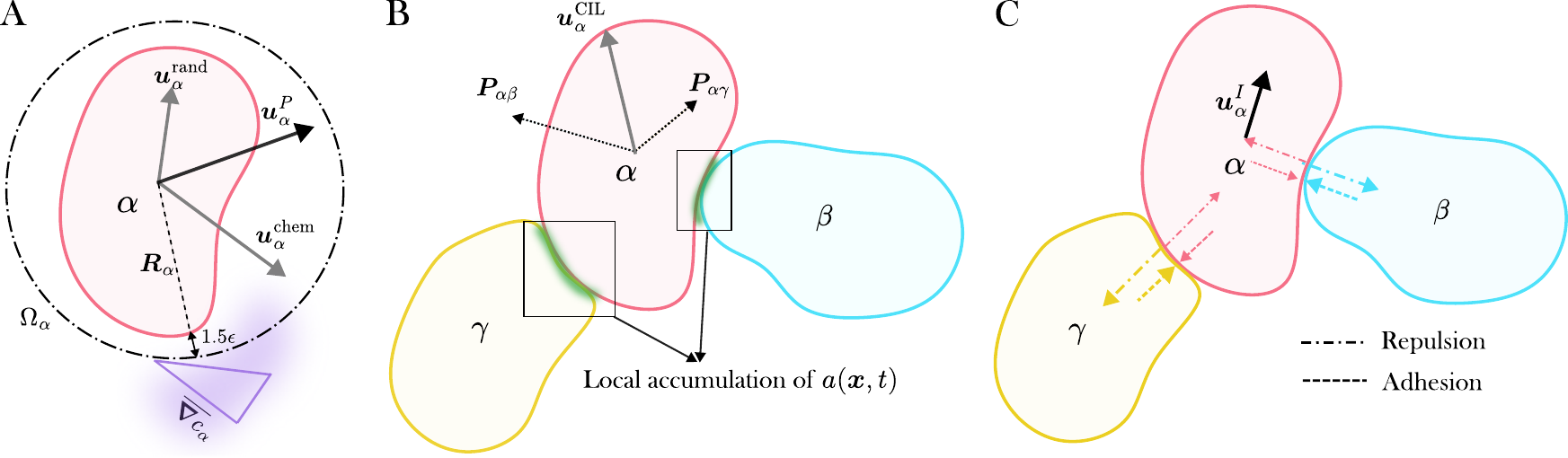}
    \caption{
    \textbf{Schematic of the polarization and interaction velocity components.}
    \textbf{(A)}
    The computational domain $\Omega_\alpha$ for cell $\alpha$ extends $1.5\epsilon$ beyond its equivalent radius $R_\alpha$ and is used to evaluate the local C3a gradient.
    In the absence of contact with another cell or a boundary,
    $\boldsymbol{u}_{\alpha}^{\mathrm{CIL}}=\boldsymbol{0}$,
    such that
    $\boldsymbol{u}_{\alpha}^{P}
    =
    \boldsymbol{u}_{\alpha}^{\mathrm{rand}}
    +
    \boldsymbol{u}_{\alpha}^{\mathrm{chem}}$.
    \textbf{(B)}
    Contact with cells $\beta$ and $\gamma$ produces local accumulation of
    $a(\boldsymbol{x},t)$
    and the pairwise polarity contributions
    $\boldsymbol{P}_{\alpha\beta}$
    and
    $\boldsymbol{P}_{\alpha\gamma}$,
    which drive the CIL response
    $\boldsymbol{u}_{\alpha}^{\mathrm{CIL}}$.
    \textbf{(C)}
    The interaction velocity
    $\boldsymbol{u}_{\alpha}^{I}$
    combines the mechanical adhesion and repulsion between cell $\alpha$ and its neighboring cells.
    }
    \label{fig:method_schematic}
\end{figure}

\noindent
We represent each neural crest cell by a phase-field $\phi_\alpha$ \cite{palmieri2015multiple,moure2021phase} that is $\sim 1$ inside the cell and $\sim 0$ outside, with a diffuse interface representing its membrane  (Fig.~\ref{fig:method_schematic}).
Cell shape evolves through advection and relaxation of a free energy that accounts for membrane surface tension, area conservation, intercellular repulsion, and intercellular adhesion (Sec.~\ref{sec:cell-evolution}).
The migration velocity of cell $\alpha$ is
$\boldsymbol{u}_{\alpha}
=
\boldsymbol{u}_{\alpha}^{P}
+
\boldsymbol{u}_{\alpha}^{I}$,
where $\boldsymbol{u}_{\alpha}^{P}$ represents active polarization component and $\boldsymbol{u}_{\alpha}^{I}$ accounts for mechanical cell-cell and cell-boundary interactions; see Eq.~\eqref{Eqn:cell_velocity_decomposition}.
The polarization velocity comprises three contributions,
$\boldsymbol{u}_{\alpha}^{P}
=
\boldsymbol{u}_{\alpha}^{\mathrm{chem}}
+
\boldsymbol{u}_{\alpha}^{\mathrm{CIL}}
+
\boldsymbol{u}_{\alpha}^{\mathrm{rand}}$,
corresponding to chemotaxis, contact inhibition of locomotion (CIL), and persistent stochastic motility, respectively.
\\

\noindent
NCCs secrete C3a, whose concentration $c(\boldsymbol{x},t)$ evolves in the extracellular space through diffusion and decay.
The C3a gradient sensed along each cell membrane determines the direction and magnitude of $\boldsymbol{u}_{\alpha}^{\mathrm{chem}}$, subject to a maximum speed, and mediates co-attraction (CoA; Sec.~\ref{sec:chemotaxis}).
We represent N-cadherin-mediated CIL through a scalar contact activity field $a(\boldsymbol{x},t)$ that captures an effective contact signal between cell-cell and cell-tissue boundaries.
This contact activity field accumulates at contact interfaces, spreads slowly along the membrane, and rapidly decays after separation; see Eq.~\eqref{Eqn:contact_activity}.
%
%
The spatial distribution of $a$ and the interface geometry determine the cell-cell and boundary polarity inputs,
$\mu\boldsymbol{P}_{\alpha}^{\mathrm{cell}}$
and
$\mu\boldsymbol{P}_{\alpha}^{\mathrm{bnd}}$,
and $\boldsymbol{u}_{\alpha}^{\mathrm{CIL}}$ relaxes towards the total CIL-driven repolarization response (see Sec.~\ref{sec:contact-inhibition-of-locomotion}).
%
%
The stochastic velocity $\boldsymbol{u}_{\alpha}^{\mathrm{rand}}$ remains fixed during each persistence interval and is updated with a directional bias toward the current polarization velocity (Sec.~\ref{sec:stochastic-velocity}).
The interaction velocity $\boldsymbol{u}_{\alpha}^{I}$ is determined by cell-cell and cell-boundary adhesion and repulsion forces derived from the free energy functional (see Sec.~\ref{sec:interaction-velocity}).
We solve the coupled field equations using the dynamic cluster field framework \cite{paspunurwar2024dynamic}, with numerical details provided in Methods.
\\

%

\subsection{\textbf{Co-attraction and cell-cell CIL fail to establish persistent directional migration of NCCs on an unconfined substrate}} \label{sec:results_no_external} 

\begin{figure}[h!]
\centering
\includegraphics[width=0.85\linewidth]{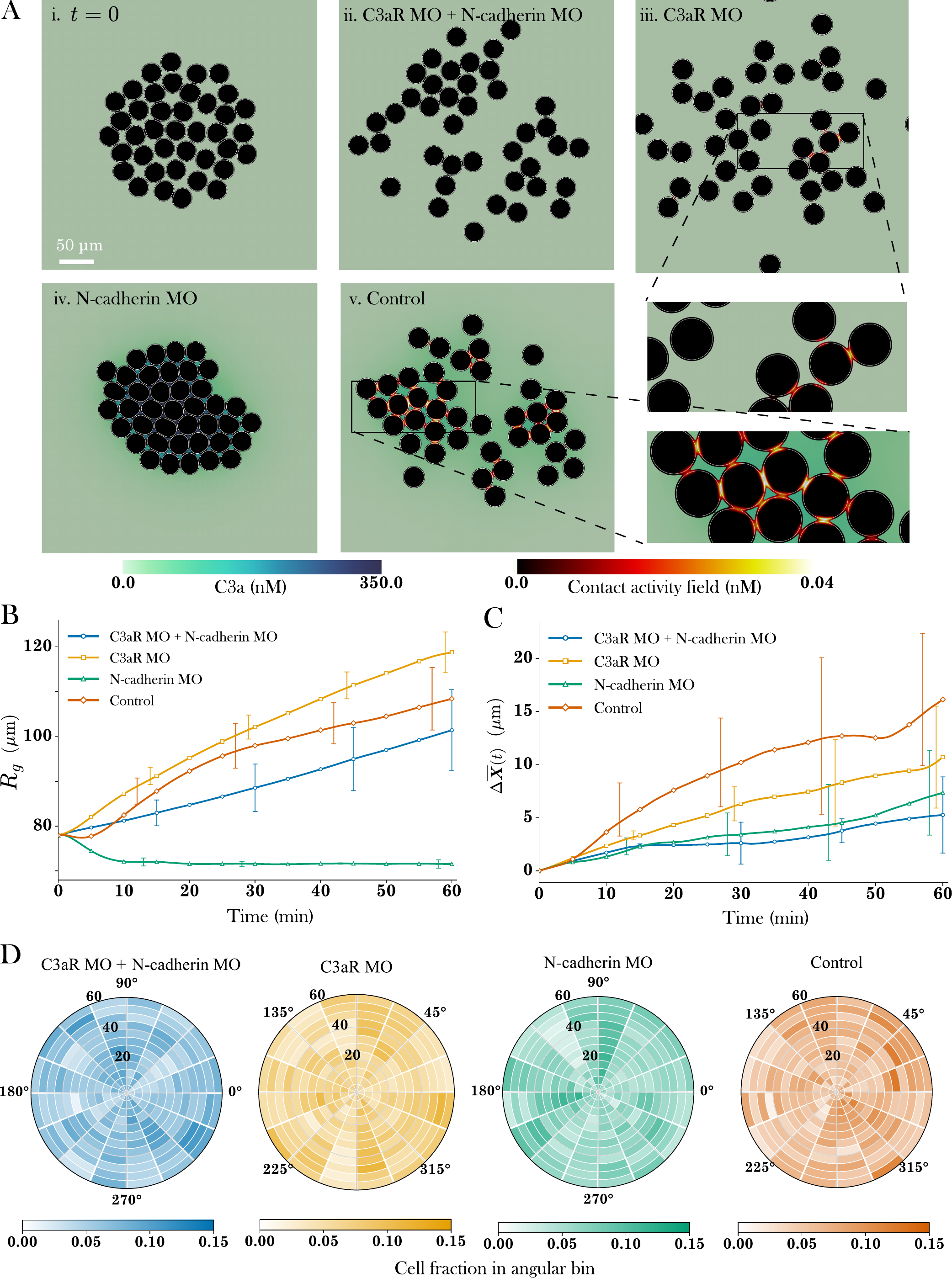}
\caption{
\textbf{Effects of C3a-mediated coattraction and cell-cell CIL on migration of NCC collectives on an unconfined substrate.}
\textbf{(A)}
Cell contours $(\phi=0.6)$, C3a concentration, and contact activity field for the common initial condition (i) and the final states of the combined C3aR and N-cadherin MO (ii), C3aR MO (iii), N-cadherin MO (iv), and control (v) conditions.
Insets enlarge contact activity at cell-cell interfaces.
Distributions of \textbf{(B)} Radius of gyration, $R_g(t)$ and \textbf{(C)} Magnitude of the population COM displacement,
$\left\lVert\Delta\overline{\boldsymbol{X}}(t)\right\rVert$, across 4 stochastic simulations for each case.
The error bars denote standard deviations.
\textbf{(D)}
Cell displacement directions over successive \SI{10}{\minute} intervals.
Each concentric band represents one interval, and color denotes the fraction of displacements within each angular bin.
}

\label{fig:case_study}
\end{figure}

\noindent We first examine migration on an unconfined planar substrate without an externally imposed chemoattractant gradient to determine whether C3a-mediated co-attraction and cell-cell CIL can maintain cohesion and establish a persistent collective direction. 
We initialize a compact collective of $N=40$ cells within a $\SI{400}{\micro\meter}\times\SI{400}{\micro\meter}$ computational domain. 
Periodic boundary conditions eliminate contact with confining boundaries and approximate migration on an extended planar substrate, i.e., $\boldsymbol{P}_{\alpha}^{\mathrm{bnd}}=\boldsymbol{0}$. 
%
%
%
We place the cells with small initial separations to establish immediate cell-cell contact and initialize the chemical and contact activity field as
$c(\boldsymbol{x},0)=0$
and
$a(\boldsymbol{x},0)=0$,
respectively
(Fig.~\ref{fig:case_study}A(i)).
To facilitate comparison with experimental perturbations, we label the simulation conditions using the corresponding morpholino terminology.
We represent C3aR MO and N-cadherin MO by setting C3a production and contact activity field production to zero, respectively.
%
Accordingly, we examine four conditions by selectively enabling C3a secretion and contact activity field production: C3aR MO + N-cadherin MO, C3aR MO, N-cadherin MO, and control.
For each condition, we simulate
$\SI{1}{\hour}$
of migration using the parameters in Table~\ref{Table:params}.
For each condition, we perform four simulations with distinct random seeds and report consistent trends across the simulations.
\\

\noindent
We use the radius of gyration as a measure of cell dispersion and collective cohesion.
The population center of mass is
\begin{equation}
\overline{\boldsymbol{X}}(t)
=
\frac{1}{N}
\sum_{\alpha=1}^{N}
\boldsymbol{X}_{\alpha}(t),
\label{Eqn:cluster_com}
\end{equation}
where $\boldsymbol{X}_{\alpha}(t)$ denotes the center of cell $\alpha$.
Unless otherwise stated, an overbar 
denotes an average over all cells. 
For cell positions, this average defines the center of mass (COM) of the NCC collective. 
The radius of gyration is
\begin{equation}
R_g(t)
=
\left[
\frac{1}{N}
\sum_{\alpha=1}^{N}
\left\lVert
\boldsymbol{X}_{\alpha}(t)
-
\overline{\boldsymbol{X}}(t)
\right\rVert^2
\right]^{1/2}.
\label{Eqn:radius_gyration}
\end{equation}
An increase in $R_g$ indicates collective expansion or dispersal, an approximately constant value indicates preservation of the collective extent, and a decrease indicates increased collective compactness.
Because the unconfined domain has no preferred migration axis, we quantify net translation of the cells using the COM displacement
\begin{equation}
\Delta\overline{\boldsymbol{X}}(t)
=
\overline{\boldsymbol{X}}(t)
-
\overline{\boldsymbol{X}}(0).
\label{Eqn:com_displacement}
\end{equation}
We assess directional coordination using rose plots of cell displacement directions over successive $\SI{10}{\minute}$ intervals, pooled across the four simulations.
A nearly uniform angular distribution indicates that the cells do not maintain a common migration direction, whereas persistent concentration within a narrow angular sector indicates coordinated directional motion.
Thus,
$R_g(t)$
measures changes in collective compactness,
$\lVert\Delta\overline{\boldsymbol{X}}(t)\rVert$
measures net population translation, and the rose plots quantify cell-level directional alignment.
Together, these measures distinguish cohesive from dispersive behavior and coordinated migration from isotropic cell motion.
\\

\paragraph{Case 1: Loss of C3a signaling and N-cadherin-mediated CIL (C3aR MO and N-cadherin MO)}

\noindent
With both co-attraction and cell-cell CIL inactive, stochastic motility independently reorients the cells, while mechanical cell-cell interactions regulate short-range adhesion and prevent overlap.
Because no chemotactic response draws outward-moving cells back toward the population, the initially compact cluster progressively disperses over \SI{1}{\hour} (Fig.~\ref{fig:case_study}A(ii)).
The radius of gyration increases from approximately \SI{80}{\micro\meter} to \SI{100}{\micro\meter} (Fig.~\ref{fig:case_study}B, blue curve with circle markers).
The COM undergoes only limited net displacement, and the displacement directions remain broadly distributed throughout the simulation (Fig.~\ref{fig:case_study}C and D, first rose plot).
Thus, stochastic motility disperses the population without organizing the cells into a common migration direction.

\paragraph{Case 2: Loss of C3a signaling (C3aR MO)}

\noindent
When C3a-mediated co-attraction is absent but cell-cell CIL remains active, contact-induced repolarization directs neighboring cells away from their shared interfaces.
Because no inward chemotactic response counteracts this separation, CIL amplifies the dispersal generated by stochastic motility.
Accordingly, $R_g$ increases more rapidly than in Case~1 (Fig.~\ref{fig:case_study}B, yellow curve with square markers).
The COM displacement also exceeds that in Case~1, but the time-resolved angular distribution remains broad and lacks a persistent preferred direction (Fig.~\ref{fig:case_study}C and D, second rose plot).
The larger COM displacement therefore reflects asymmetric dispersal within a noncohesive population rather than coordinated collective migration.
The difference between Cases~1 and~2 isolates the dispersive effect of cell-cell CIL when it acts without co-attraction.
This behavior agrees qualitatively with experiments showing that disruption of C3a signaling reduces neural crest cohesion and promotes dispersal \cite{theveneau2010collective,mclennan2010vascular}.

\paragraph{Case 3: Loss of N-cadherin-mediated CIL (N-cadherin MO)}

\noindent
Without cell-cell CIL, $R_g$ decreases as continued C3a secretion generates inward-directed chemotactic responses and compacts the collective (Fig.~\ref{fig:case_study}B, green curve).
The cells therefore remain cohesive and form a densely packed aggregate with pronounced deformation at cell-cell interfaces (Fig.~\ref{fig:case_study}A(iv)).
The lower steady value of $R_g$ relative to cases 1 and 2 shows that co-attraction promotes compaction when CIL does not oppose close packing.
Despite this cohesion, the COM undergoes little net translation, and the displacement directions remain broadly distributed over successive time intervals (Fig.~\ref{fig:case_study}C and D, third rose plot).
Thus, C3a-mediated co-attraction maintains cohesion and promotes compaction but does not independently establish population-wide directional migration.

\paragraph{Case 4: Intact C3a and N-cadherin signaling (control)}

With both mechanisms active, Cell-cell CIL repolarizes contacting cells away from their shared interfaces, while CoA draws nearby cells together and promotes local clustering.
The control exhibits the largest mean COM displacement among the four conditions, reaching approximately \SI{16}{\micro\meter} over \SI{1}{\hour}, equivalent to about one cell radius (Fig.~\ref{fig:case_study}C).
Although $R_g$ exceeds that in Case~1, CoA promotes local aggregation of dispersed cells into smaller clusters rather than preserving a single compact aggregate (Fig.~\ref{fig:case_study}A(v)).
The larger collective extent therefore coexists with local cohesion.
The control exhibits the largest mean COM displacement among the four conditions, reaching approximately \SI{16}{\micro\meter} over \SI{1}{\hour} (Fig.~\ref{fig:case_study}C).
However, the angular distribution shows only transient directional biases, with no common direction persisting across successive time intervals (Fig.~\ref{fig:case_study}D, control).
Thus, CoA promotes local clustering despite CIL-mediated separation, but the combined interactions do not establish persistent directional migration on an unconfined substrate.
\\

\noindent
Together, the four conditions distinguish the roles of CoA and cell-cell CIL in collective organization.
Without CoA, cells disperse, and retaining cell-cell CIL increases the dispersal rate, as shown by the comparison between Cases~1 and~2.
CoA alone produces the most compact aggregate, whereas the control exhibits greater spatial spread while retaining local clustering.
Despite these differences in cellular organization, none of the four conditions establishes a persistent collective migration direction on an unconfined substrate without external chemotactic guidance.
\\




\subsection{\textbf{Boundary-contact repolarization enables directional migration of NCC collectives under confinement}}
\label{sec:confined-migration}

\noindent
The preceding results show that C3a-mediated co-attraction maintains collective cohesion, whereas cell-cell CIL produces local contact-induced repolarization.
Neither mechanism consistently establishes a persistent collective directional bias in the absence of spatial confinement.
However, in vivo, NCCs migrate through spatially confined streams, with nonpermissive regions in the surrounding tissues restricting their migratory paths \cite{szabo2016vivo}.
Directional collective migration is also observed in vitro within restricted spaces and on fibronectin channels \cite{carmona2011complement} bordered by versican \cite{szabo2016vivo}.
We therefore ask whether confinement acts only as a passive geometric restriction or whether contact with the confining boundary provides an active polarity cue that supports outlet-directed migration.
We further examine how collective size and the transverse freedom permitted by a widened corridor regulate the establishment and persistence of this directional response.
\\

\noindent
We consider a rectangular corridor
$\Omega=(0,L_x)\times(0,L_y)$
with a closed rear boundary at $x=0$, lateral boundaries at $y=0$ and $y=L_y$, and an open outlet at $x=L_x$.
We refer to the $x$ direction as the axial direction, with positive $x$ pointing toward the open outlet, and the $y$ direction as the lateral direction.
Unless otherwise stated, we initialize the cells in compact arrangements adjacent to the rear boundary and spanning the corridor width.
We choose $L_x$ such that the initial axial distance between the cell nearest the outlet and the outlet is $\SI{320}{\micro\meter}$, and initialize $c(\boldsymbol{x},0)=0$ and
$a(\boldsymbol{x},0)=0$, so that no pre-existing chemical gradient or contact activity field biases the initial migration direction.
\\

\noindent We first examine a two-cell pair as the smallest collective capable of exhibiting contact-mediated migration. 
With boundary-contact repolarization active, the pair establishes sustained outlet-directed migration. 
To isolate the role of this response, we set $\boldsymbol{P}_{\alpha}^{\mathrm{bnd}}=\boldsymbol{0}$ while preserving the corridor geometry and mechanical cell-boundary adhesion and repulsion. 
Under this matched control, the pair fails to establish a sustained outlet-directed bias. 
%
%
%
%
%
We next extend the analysis to collectives of
$N\in\{4,9,16,25,36\}$
cells to determine how collective size affects the establishment and maintenance of directional migration.
For each collective size, we set the corridor height to
$L_y=\sqrt{N}D$,
where $D$ is the initial cell diameter, so that the initial cell arrangement spans the corridor width.
%
%
Finally, for the $16$-cell collective, we introduce a
$\SI{30}{\micro\meter}$
gap between each outer cell row and the corresponding lateral boundary.
This configuration tests how increased transverse freedom and reduced lateral boundary contact affect collective directionality.
\\

\subsubsection{\textbf{Boundary-contact repolarization initiates and sustains robust two-cell migration under tight confinement}}
\label{sec:two-cell-confinement}

\begin{figure}[!t]
\centering
\includegraphics[width=\linewidth]{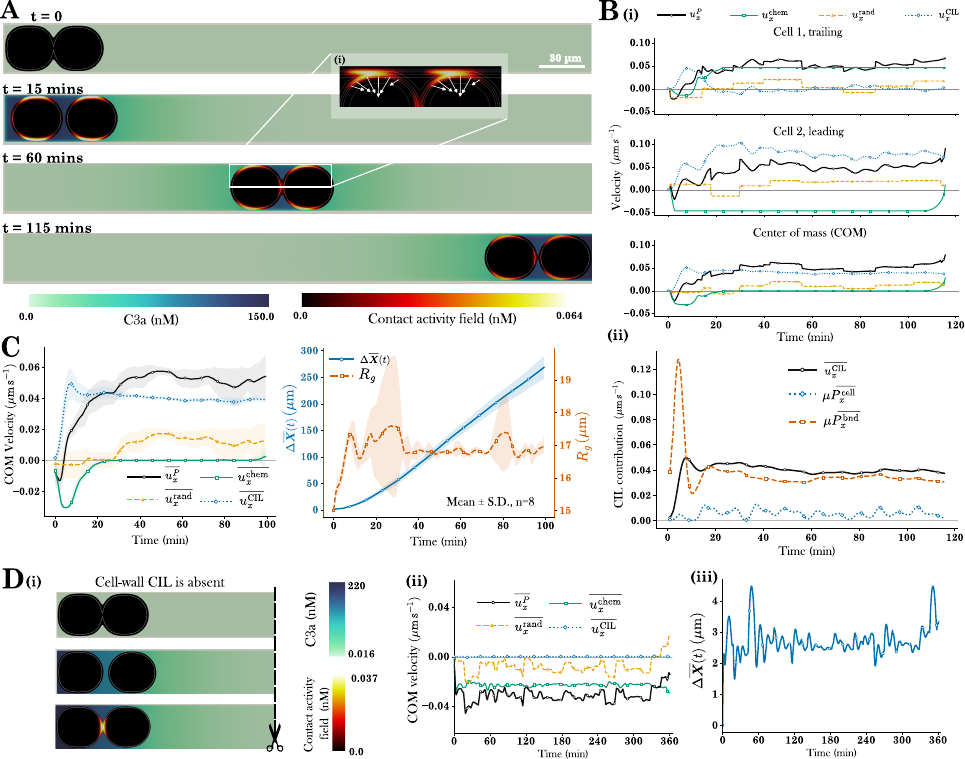}
\caption{
\textbf{Boundary-contact repolarization initiates and sustains robust two-cell migration under tight confinement.}
\textbf{(A)}
Representative cell, C3a, and contact activity field snapshots at $0$, $15$, $60$, and $\SI{115}{\minute}$ during migration toward the open outlet. The inset (i) shows the local boundary-contact polarity contribution.
\textbf{(B)} (i) Axial polarization velocity and its chemotactic, stochastic, and CIL components for the trailing cell (Cell 1), leading cell (Cell 2), and Center of mass (COM).
(ii) Axial CIL velocity at COM, $\overline{u_{x}^{\mathrm{CIL}}}$, and the axial cell-cell and boundary-contact polarity response, $\mu\overline{P_{x}^{\mathrm{cell}}}$ and $\mu\overline{P_{x}^{\mathrm{bnd}}}$, respectively.
\textbf{(C)} Distribution of the COM polarization velocity components, COM displacement, and radius of gyration $R_g$ across eight simulations.
\textbf{(D)}(i) The cell, C3a, and contact activity field states, (ii) Axial polarization velocity and its components of the COM, and (iii) axial COM displacement for matched control with $\boldsymbol{P}_{\alpha}^{\mathrm{bnd}}=\boldsymbol{0}$.
}
\label{fig:corridor1}
\end{figure}

\noindent
In the one-cell-wide corridor $(L_y=D)$, the cells partially partition the confined domain, causing the self-generated C3a field to accumulate between the initially unpolarized pair and the closed rear boundary (Fig.~\ref{fig:corridor1}A).
This concentration asymmetry produces a rear-directed COM chemotactic contribution, and the pair initially moves toward the closed boundary (Fig.~\ref{fig:corridor1}B(i)).
As the rear cell contacts the rear boundary, the contact activity field accumulates at the cell-boundary interface, generating an outlet-directed boundary-polarity input.
The resulting CIL response progressively opposes the rear-directed chemotactic contribution.
Once the CIL contribution becomes sufficiently strong, the COM polarization velocity changes sign, and the pair reverses its migration, moving toward the outlet.
\\

\noindent
After reversal, the cells migrate together toward the outlet as a cohesive pair, while the C3a field remains centered on the pair and extends along the corridor (Fig.~\ref{fig:corridor1}A).
CoA pulls the trailing cell forward and the leading cell backward toward their shared C3a-rich region, causing their opposing chemotactic velocities to produce little net contribution to COM motion (Fig.~\ref{fig:corridor1}B(i)).
The stochastic contribution fluctuates in time but develops a smaller positive mean after reversal, thereby reinforcing rather than initiating outlet-directed migration.
In contrast, the CIL velocity remains the dominant positive contribution to the COM polarization velocity, thereby sustaining motion toward the outlet.
The COM velocity balance shows that CoA maintains pair cohesion and stochastic motility reinforces the selected direction, whereas the CIL response provides the principal outlet-directed contribution (Fig.~\ref{fig:corridor1}B(i), bottom).
\\

\noindent
To identify the source of the CIL response, we compare the cell-averaged axial CIL velocity,
$\overline{u_{x}^{\mathrm{CIL}}}$,
with the cell-averaged axial cell-cell and boundary-contact polarity inputs,
$\mu\overline{P_{x}^{\mathrm{cell}}}$
and
$\mu\overline{P_{x}^{\mathrm{bnd}}}$,
respectively.
When the pair contacts the closed rear boundary, the polarity contribution from that boundary drives
$\mu\overline{P_{x}^{\mathrm{bnd}}}$
strongly positive, generating the outlet-directed CIL response that reverses the pair's initial rear-directed motion (Fig.~\ref{fig:corridor1}B(ii)).
After the pair moves away from the rear boundary, the boundary-contact input $\mu\overline{P_{x}^{\mathrm{bnd}}}$ remains predominantly positive and continues to provide the principal outlet-directed input to
$\overline{u_{x}^{\mathrm{CIL}}}$.
In contrast,
$\mu\overline{P_{x}^{\mathrm{cell}}}$
remains comparatively small and does not generate a sustained axial bias in the COM CIL response.
Thus, boundary-contact repolarization both initiates the reversal and sustains the subsequent outlet-directed migration, whereas cell-cell CIL remains locally active without providing the dominant collective directional cue.
\\

\noindent Because the cell membrane is curved along the boundary-contact region, the local contact-induced polarity directions contain both lateral and axial components.
As the pair advances, the leading side of each cell continually establishes new boundary contact, where contact activity field $a$ has not yet accumulated, whereas older contact regions near the trailing side retain larger values of $a$. 
This contact history produces an asymmetric distribution of $a$ along the curved cell-boundary interface (Fig.~\ref{fig:corridor1}A, inset (i)). 
For a symmetric distribution, the opposing axial components of the local interface normals would cancel. 
The asymmetric weighting prevents this cancellation and produces a nonzero outlet-directed component of $\boldsymbol{P}_{\alpha}^{\mathrm{bnd}}$. 
Consequently, continued contact with the lateral boundaries sustains a positive $\mu\overline{P_{x}^{\mathrm{bnd}}}$ after the pair moves away from the rear boundary.
\\

\noindent
This velocity balance persists across all eight stochastic simulations (Fig.~\ref{fig:corridor1}C).
The mean CIL velocity of the COM remains positive across the simulations, whereas the COM chemotactic contribution approaches zero after the initial transient, and the stochastic contribution maintains a smaller positive mean.
Consequently, the COM displacement increases steadily with a narrow simulation-to-simulation spread.
At the same time, $R_g$ remains stable throughout migration (see Fig.~\ref{fig:corridor1}C).
Since $R_g(t) \approx R$ for all times, pair cohesion is maintained throughout sustained migration.
\\

\noindent
To isolate active boundary guidance from mechanical confinement, we perform a matched control in which we suppress the boundary polarity contribution by setting
$\boldsymbol{P}_{\alpha}^{\mathrm{bnd}}=\boldsymbol{0}$.
The corridor geometry, mechanical cell-boundary interactions, cell-cell CIL, CoA, and stochastic motility remain unchanged (Fig.~\ref{fig:corridor1}D).
Without the boundary polarity input, the chemotactic contribution remains rear-directed, while cell-cell CIL contribution provides essentially no net COM motion.
Stochastic fluctuations do not establish a sustained outlet-directed polarization velocity (Fig.~\ref{fig:corridor1}D(ii)).
The pair consequently moves only a few micrometers over $\SI{6}{\hour}$ and remains near the closed rear boundary (Fig.~\ref{fig:corridor1}D(i) and D(iii)).
\\

\noindent
Together, these comparisons identify boundary-contact repolarization as the directional cue required for sustained outlet-directed migration under the tested conditions.
CoA maintains pair cohesion, while stochastic motility can reinforce the direction established by boundary contact but does not provide a persistent outlet-directed bias in its absence.
The confining boundaries therefore support migration by actively reorienting cell polarity, rather than through spatial restriction alone.
\\

\subsubsection{\textbf{Collective size modulates robustness of directional persistence under confinement}}
\label{sec:collective-size-polarity}

\begin{figure}[!t]
\centering
\includegraphics[width=\linewidth]{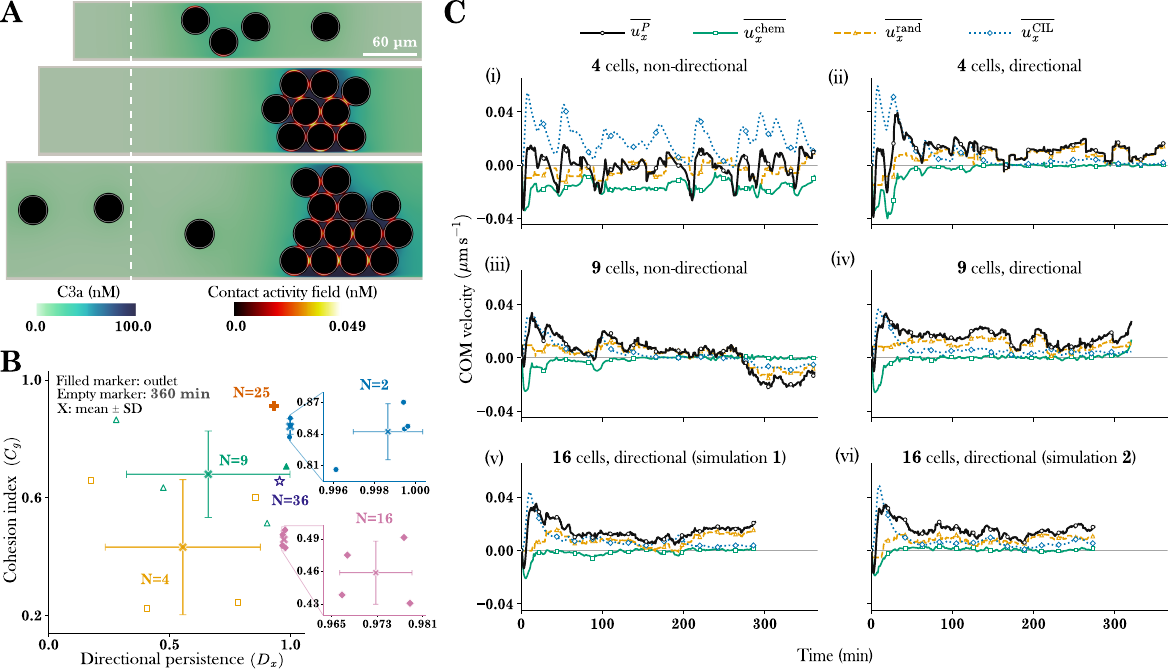}
\caption{
\textbf{Collective size regulates directional robustness through polarity coordination under tight confinement.}
\textbf{(A)} Representative cell configurations and corresponding C3a and contact activity field for collectives containing $4$, $9$, and $16$ cells. 
\textbf{(B)} Signed axial directional persistence, $\mathcal{D}_x$, versus cohesion index, $C_g$, for $N\in\{2,4,9,16,25,36\}$.
Symbols denote individual simulations, with filled symbols indicating arrival at the outlet and empty symbols indicating the state at $360~\mathrm{min}$. 
Crosses denote the mean across multiple simulations, and the insets enlarge the $2$-cell and $16$-cell results. 
\textbf{(C)} COM polarization velocity and its chemotactic, stochastic, and CIL components for nondirectional and directional $4$-cell and $9$-cell simulations and two directional $16$-cell simulations. 
}
\label{fig:corridor2}
\end{figure}

\noindent
Collective size changes the number and spatial distribution of cell-cell and cell-boundary contacts under tight confinement (Fig.~\ref{fig:corridor2}A).
Representative $4$-, $9$-, and $16$-cell collectives remain laterally confined, with the self-generated C3a field centered on each collective as it migrates and contact activity concentrated at cell-cell and cell-boundary interfaces.
As the collective size increases, the number of cell rows and internal cell-cell interfaces increases, while cells along the outer rows remain in contact with the lateral boundaries.
Collective size, therefore, changes how local CIL and stochastic polarity contributions combine at the population level, without removing the boundary-contact cue.
\\

\noindent
To distinguish directed translation from preservation of collective compactness, we evaluate each simulation at the final time $t_f$ using the signed axial directional persistence $\mathcal{D}_x$ and the cohesion index $C_g$, defined as
\begin{align}
\mathcal{D}_x
&=
\frac{
\overline{X}_x(t_f)-\overline{X}_x(0)
}{
\displaystyle
\int_{0}^{t_f}
\left\lVert
\frac{\mathrm{d}\overline{\boldsymbol{X}}}{\mathrm{d}t}
\right\rVert
\,\mathrm{d}t
},
\label{Eqn:directional_persistence}
\\
C_g
&=
\frac{R_g(0)}{R_g(t_f)}.
\label{Eqn:cohesion_index}
\end{align}
Here, $\overline{\boldsymbol{X}}(t)$ is the COM of the cell collective defined in Eq.~\eqref{Eqn:cluster_com}, and $\overline{X}_x(t)$ denotes its axial coordinate.
The numerator of $\mathcal{D}_x$ measures signed axial displacement of the COM, whereas the denominator measures total distance traveled by the COM.
A value of $\mathcal{D}_x=1$ corresponds to a straight outlet-directed trajectory, $\mathcal{D}_x\approx0$ indicates negligible net axial displacement relative to the total distance traveled, and $\mathcal{D}_x<0$ indicates net motion toward the rear boundary.
A value of $C_g=1$ indicates that the collective retains its initial spatial extent, whereas $C_g>1$ indicates compaction and $C_g<1$ indicates expansion.
We plot $\mathcal{D}_x$ against $C_g$ for $N\in\{2,4,9,16,25,36\}$ in Fig.~\ref{fig:corridor2}B.
The two metrics vary independently, indicating that the preservation of collective compactness does not, by itself, determine directional migration. 
For $N=2$, the simulations combine near-unity directional persistence with relatively preserved pair spacing. 
For $N=16$, the simulations also maintain $\mathcal{D}_x\approx 1$, but exhibit substantially lower values of $C_g$. 
The representative $16$-cell configuration contains several cells separated from the main aggregate, which increases $R_g$ without disrupting the outlet-directed motion of the population COM (Fig.~\ref{fig:corridor2}A, bottom). 
Thus, high directional persistence can co-exist with partial loss of cohesion. 
\\

\noindent The $4$-cell and $9$-cell simulations span a broader region of the $\mathcal{D}_x$-$C_g$ state space. 
Some simulations remain relatively cohesive but exhibit weak directional persistence, whereas others maintain an outlet-directed COM trajectory despite considerable collective expansion. 
Cohesion is therefore neither sufficient to guarantee directionality nor strictly required for the population COM to migrate persistently toward the outlet. 
Instead, the migration outcome depends on whether the cell-level polarity contributions combine into a sustained positive COM polarization velocity. 
\\

\noindent The contrast between the failed (non-directional) and successful (directional) $4$-cell simulations illustrates the role of stochastic motility in this coordination process (Fig.~\ref{fig:corridor2}C(i) and C(ii)). %
Both simulations initially exhibit a positive $\overline{u_{x}^{\mathrm{CIL}}}$ that opposes a rear-directed $\overline{u_{x}^{\mathrm{chem}}}$. %
The rear-directed chemotactic contribution arises because C3a accumulates in the confined region between the collective and the closed rear boundary, producing a concentration gradient toward the closed end. 
In the successful simulation, $\overline{u_{x}^{\mathrm{rand}}}$ aligns with the CIL response, producing a positive $\overline{u_{x}^{P}}$ and allowing the collective to move away from the rear boundary. 
In the failed simulation, the stochastic contribution fluctuates relative to the CIL response, while the opposing chemotactic contribution keeps the net polarization velocity near zero. 
Consequently, the collective remains near the closed end despite retaining an outlet-directed CIL contribution. The stochastic contribution, therefore, does more than merely add unstructured fluctuations to the COM velocity. 
As described in Sec.~\ref{sec:stochastic-velocity}, updated stochastic directions are biased toward the existing polarization velocity through $\rho_{\mathrm{pol}}$. 
An initial boundary-induced CIL response can therefore bias subsequent stochastic motility toward the outlet, creating a positive feedback that reinforces the selected direction. 
Conversely, stochastic motion that initially opposes this response can weaken the outlet-directed bias and prevent the collective from escaping the rear boundary. 
This competition explains why simulations with the same collective size and parameters can produce different migration outcomes.
\\

\noindent The $9$-cell simulations reveal a second mode of directional failure (Fig.~\ref{fig:corridor2}C(iii) and C(iv)). 
The failed simulation initially develops a positive COM polarization velocity and moves toward the outlet.
At later times, however, the CIL and stochastic contributions lose their positive alignment, causing $\overline{u_{x}^{P}}$ to weaken and eventually become rear-directed. 
The collective therefore reverses its migration direction after initially establishing outlet-directed motion. 
In contrast, the successful $9$-cell simulation maintains a positive combined CIL and stochastic contribution throughout most of the trajectory. 
Intermediate collective sizes can therefore fail either by remaining trapped near the rear boundary or by losing polarity coordination after migration has begun. 
CoA can preserve collective cohesion in both cases, but it does not restore an outlet-directed polarity bias. 
\\

\noindent
After the initial transient, the two $16$-cell simulations maintain a positive
$\overline{u_{x}^{P}}$
despite different temporal fluctuations in the underlying velocity components (Fig.~\ref{fig:corridor2}C(v) and C(vi)).
They also sustain robust outlet-directed migration despite the partial separation of several cells from the main collective, as shown in Fig.~\ref{fig:corridor2}A (bottom).
Thus, directional persistence does not require that every cell remains within a single compact aggregate or that the velocity components follow the same temporal evolution.
Instead, the boundary response is distributed among several cells, so local polarity loss or separation of a small subset does not determine the direction of the population COM. 
Averaging over a larger number of cells also reduces the influence of uncoordinated stochastic fluctuations, while stochastic contributions biased toward the established polarization reinforce the outlet-directed state. 
We extended the analysis to $N=25$ and $N=36$, and the single simulation performed at each size similarly maintained outlet-directed migration as seen in Fig.~\ref{fig:corridor2}B, qualitatively supporting the directional robustness observed for larger collectives. 
\\

\noindent These results suggest two distinct routes to robust migration. 
The two-cell pair relies on a simple leader-trailer organization established by rear-boundary repolarization, whereas the $16$-cell collective achieves robustness through distributed boundary sensing and collective-level averaging. 
The intermediate $4$-cell and $9$-cell collectives have more internal degrees of freedom than the pair but less redundancy than the larger collective. 
They are therefore more susceptible to stochastic polarity configurations that either prevent initial escape or cause a later reversal. 
Collective size consequently regulates directional robustness by enabling the population to reinforce and maintain the boundary-induced polarity cue, rather than through cohesion alone. 
\\

\subsubsection{\textbf{Persistent contact with a lateral boundary sustains directional migration in a widened corridor}}
\label{sec:transverse-freedom-detachment}

\begin{figure}[!ht]
\centering
\includegraphics[width=\linewidth]{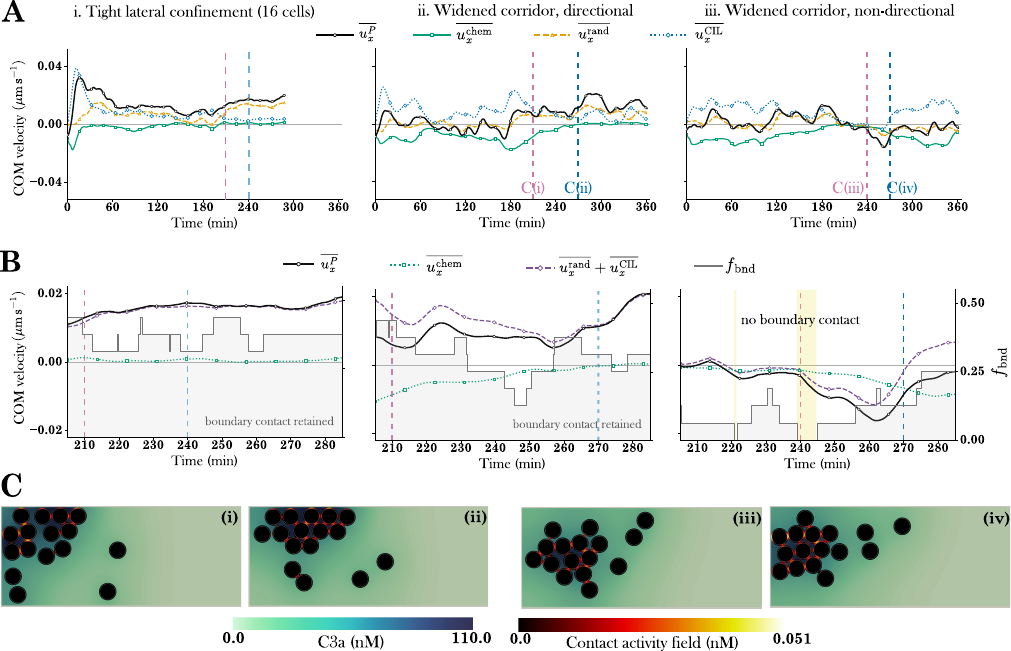}
\caption{
\textbf{Lateral boundary association sustains directional migration when confinement is relaxed.}
\textbf{(A)}
Axial COM polarization velocity,
$\overline{u_{x}^{P}}$,
and its chemotactic, stochastic, and CIL components for
(i) tight lateral confinement,
(ii) directional migration in a widened corridor,
and
(iii) nondirectional migration in the same widened corridor.
The widened corridor includes a $\SI{30}{\micro\meter}$ initial gap between the collective and each lateral boundary.
Dashed vertical lines mark the snapshot times shown in panel C.
\textbf{(B)}
Magnified axial COM polarization velocity, chemotactic contribution, combined stochastic and CIL contribution, and
$f_{\mathrm{bnd}}$,
from $205$ to $285~\mathrm{min}$.
Yellow shading denotes the interval without lateral boundary contact.
\textbf{(C)}
Cell configurations and corresponding C3a and contact activity field at the marked times for the directional
(i, ii)
and nondirectional
(iii, iv)
simulations in the widened corridor.
}
\label{fig:corridor3}
\end{figure}

\noindent
We next ask whether sustained directional migration requires the collective to remain tightly confined between both lateral boundaries.
To test this, we introduce a $\SI{30}{\micro\meter}$ gap between the outer cell rows and each lateral boundary, allowing the collective to move laterally, detached completely from both boundaries.
The simulations produce two distinct outcomes in the same widened corridor (Fig.~\ref{fig:corridor3}).
Compared with the tightly confined reference, one collective remains associated with the upper lateral boundary and continues migrating toward the outlet, whereas the other detaches from both lateral boundaries and stalls despite retaining its self-generated C3a field and cell-cell contacts.
Thus, tight bilateral confinement is not required, but sustained migration remains associated with continued contact with at least one lateral boundary.
\\

\noindent The axial COM velocity components show that sustained migration depends on whether the collective escapes the chemotactic bias near the closed rear boundary. 
In all three simulations, C3a initially accumulates between the collective and the rear boundary, producing a negative $\overline{u_{x}^{\mathrm{chem}}}$, while rear boundary contact generates an opposing positive CIL response. 
Under tight lateral confinement, continued contact with both lateral boundaries sustains this response and, together with stochastic polarity, moves the collective away from the closed end (Fig.~\ref{fig:corridor3}A(i)). 
The directional simulation in the widened corridor follows the same sequence, with association with the upper lateral boundary sustaining the outlet-directed response (Fig.~\ref{fig:corridor3}A(ii)). 
%
Once the collective advances sufficiently, continued C3a secretion shifts the concentration field with the moving cells, producing a more symmetric C3a distribution around the collective and causing
$\overline{u_{x}^{\mathrm{chem}}}$
to approach zero. 
\\

\noindent The nondirectional simulation in the widened corridor does not complete this transition (Fig.~\ref{fig:corridor3}A(iii)). 
Without sustained association with either lateral boundary, the initial positive response does not develop into a stable outlet-directed velocity balance. 
The collective remains near the closed end, where $\overline{u_{x}^{\mathrm{chem}}}$ stays negative and the CIL and stochastic contributions only intermittently overcome it. 
Consequently, $\overline{u_{x}^{P}}$ fluctuates near zero or becomes negative, and the collective repeatedly moves toward and away from the rear boundary. 
Thus, rear boundary contact initiates motion toward the outlet, whereas lateral boundary association enables the collective to escape the rear chemotactic bias and establish sustained migration. 
\\

\noindent
To relate lateral boundary contact to the axial polarity balance, we define the fraction of cells contacting either lateral boundary as
\begin{equation}
f_{\mathrm{bnd}}(t)
=
\frac{N_{\mathrm{bnd}}(t)}{N},
\label{Eqn:boundary_contact_fraction}
\end{equation}
where $N_{\mathrm{bnd}}(t)$ denotes the number of cells in contact with a lateral boundary at time $t$.
In the tightly confined and directional widened corridor simulations,
$f_{\mathrm{bnd}}$
remains nonzero, while
$\overline{u_{x}^{\mathrm{rand}}}
+
\overline{u_{x}^{\mathrm{CIL}}}$
remains predominantly positive
(Fig.~\ref{fig:corridor3}B(i) and B(ii)).
The snapshots in Fig.~\ref{fig:corridor3}C(i) and C(ii) confirm that the directional collective remains associated with the upper lateral boundary as it advances for the directional widened corridor simulation.
In the nondirectional simulation, the interval with
$f_{\mathrm{bnd}}=0$
coincides with the combined CIL and stochastic contribution becoming negative and with a decrease in
$\overline{u_{x}^{P}}$
(Fig.~\ref{fig:corridor3}B(iii) and C(iii)).
After the collective recontacts a lateral boundary, the positive axial balance recovers only gradually, and sustained migration is not immediately restored
(Fig.~\ref{fig:corridor3}B(iii) and C(iv)).
This delay is consistent with the time required for contact activity to rebuild at the boundary and for the CIL and stochastic contributions to realign toward the outlet.
Together with the boundary polarity ablation in Fig.~\ref{fig:corridor1}D, these results support distinct roles for the confining boundaries: rear boundary contact initiates the outlet-directed response, whereas lateral boundary contact sustains it.
\\

\noindent
These comparisons show that sustained directional migration does not require simultaneous contact with both lateral boundaries.
Cell-boundary CIL at the rear boundary initiates outlet-directed motion, while cell-boundary CIL generated through association with at least one lateral boundary sustains the polarity response needed to escape the C3a-rich region near the closed end.
After escape, the chemotactic contribution approaches zero, and the boundary-induced CIL and stochastic contributions maintain migration toward the outlet.
Complete detachment removes this cell-boundary CIL input, allowing the persistent rear-directed chemotactic bias to stall or reverse the collective.
Thus, widening the corridor affects migration through the boundary contact states it permits, showing that confining boundaries actively regulate cell polarity rather than acting solely as passive geometric restrictions.
\\


\section{Discussion}

\noindent
Understanding collective NCC migration requires distinguishing the mechanisms that preserve cohesion from those that sustain a common direction.
Previous studies establish roles for CoA, CIL, polarity persistence, and confinement, but their coupled action makes the individual contributions difficult to isolate \cite{woods2014directional,szabo2016vivo,merchant2018rho,merchant2020rho}.
The central question is how boundary contact contributes to directional guidance beyond mechanical confinement, particularly without an external chemoattractant gradient.
Our framework builds on DCF \cite{paspunurwar2024dynamic} and resolves C3a transport in the evolving extracellular space, allowing chemical gradients to respond to cell arrangement and confinement.
We represent contact-induced repolarization through a shared contact activity field, separately from mechanical adhesion and repulsion.
This separation allows us to suppress the boundary polarity response while preserving the corridor geometry and mechanical interactions.
\\

%
%
%

\noindent After constraining the relevant parameters using experimental pairwise collision dynamics \cite{scarpa2013novel}, we examine NCC migration in an unconfined substrate.
The simulations reveal that C3a-mediated CoA promotes cell association, while N-cadherin-mediated cell-cell CIL regulates spacing and rearrangement, but neither establishes persistent directional migration.
In contrast, under confinement, C3a initially attracts the cells toward the rear boundary, where contact-induced repolarization reverses their polarization velocity toward the outlet.
As outlet-directed migration proceeds, asymmetric contact activity along the curved lateral cell-boundary interfaces produces an axial polarity contribution that helps sustain outlet-directed motion.
Suppressing the polarity response from boundary contact, while retaining the corridor geometry and mechanical cell-boundary interactions, prevents sustained outlet-directed migration.
Spatial restriction alone is therefore insufficient to sustain directional migration under the tested conditions.
These findings complement previous models of prescribed boundary guidance and intracellular polarity persistence \cite{szabo2016vivo,merchant2018rho,merchant2020rho} by distinguishing active boundary repolarization from mechanical confinement and resolving its competition with self-generated chemotaxis.
\\

\noindent
Comparisons across collective sizes and with a widened corridor show how robustly this boundary-contact cue is maintained.
In the $4$-cell and $9$-cell simulations, stochastic polarity can reinforce or oppose the CIL response, causing the collective to migrate, remain trapped near the closed boundary, or reverse its migration after it begins.
The $16$-cell simulations remain directional despite partial cell separation, consistent with a boundary response distributed among more cells and a smaller influence of isolated polarity fluctuations on the COM motion.
In the widened corridor, migration persists while the collective remains associated with one lateral boundary, whereas complete detachment accompanies loss of directional persistence.
Together, these results link directional robustness to continued boundary association and coordination of the polarity contributions, rather than collective compactness or bilateral confinement alone.
\\


\noindent
The model remains a reduced representation of NCC migration.
%
%
The model does not explicitly resolve Rac1 and RhoA dynamics, cell-matrix mechanics, matrix remodeling, or distinct leader and follower states \cite{trepat2009physical,wynn2012computational,kabla2012collective}.
It also omits additional intracellular and collective signaling mechanisms considered in other models \cite{camley2016emergent,camley2017physical}.
%
%
The predicted response to boundary contact should therefore be interpreted as a mechanistic hypothesis rather than an established molecular pathway.
This hypothesis yields direct experimental predictions.
Experiments could compare NCC migration in corridors with identical geometry but different boundary coatings.
Imaging protrusions and Rac1 or RhoA activity during and after contact could test whether boundary encounters induce local repolarization and whether it persists after separation.
Associated changes in directional persistence would support active boundary guidance, provided that differences in adhesion or baseline motility do not account for them.
Experiments in widened corridors could further determine whether contact with a single lateral boundary sustains migration and whether complete detachment precedes stalling or reversal.
\\

\noindent
Future directions can assign distinct response kinetics to cell-cell and cell-boundary contact and incorporate explicit intracellular signaling, cell-matrix interactions, and tissue geometry in three dimensions.
The model can also represent differences among cells at the leading, interior, and trailing regions of a collective without assigning these identities in advance.
Including placode cells and placode-derived Sdf1 would enable the study of chase-and-run migration \cite{theveneau2013chase,colombi2020modelling}, with Sdf1 transported dynamically through the extracellular space in the same manner as C3a.
These developments would connect the predicted boundary response to specific molecular and tissue mechanisms.
More broadly, the present framework provides a basis for testing how self-generated chemotaxis and contact with confining tissues jointly regulate collective guidance during development and invasive migration.
\\

\clearpage


\section{Methods}

\subsection{\textbf{Cell evolution dynamics}}\label{sec:cell-evolution}

\noindent
We use a two-dimensional phase-field model \cite{moure2021phase,anderson1989colloid,gomez2019review} to study multicellular migration of \textit{Xenopus} NCCs on a planar substrate, driven by chemotaxis and contact inhibition of locomotion.
Let $\Omega$ denote the computational domain.
The system comprises of $N$ cells, with cell $\alpha$ represented by the phase field $\phi_\alpha(\boldsymbol{x},t)$.
The phase field captures the cell location and shape, taking values of $\sim 1$ inside the cell and $\sim 0$ outside, with a diffuse interface of thickness $\epsilon$.
The evolution of this multicellular system \cite{palmieri2015multiple} is governed by the energy functional
\begin{equation}
\mathcal{F}[\phi_1,\ldots,\phi_N]
=
\sum_{\alpha=1}^{N}
\mathcal{F}_{\alpha}[\phi_1,\ldots,\phi_N],
\label{Eqn:total-energy-functional}
\end{equation}
where the $\alpha$-cell energy functional $\mathcal{F}_{\alpha}$ accounts for membrane surface tension, cell-area conservation, cell-cell repulsion, and cell-cell adhesion:
\begin{eqnarray} 
    \mathcal{F_\alpha}[\phi_1,...,\phi_N] 
    &=& \int_\Omega\gamma\left(\frac{G(\phi_\alpha)}{\epsilon^2} 
    + \abs{\nabla\phi_\alpha}^2 \right)\,{\rm d}\Omega
    +\frac{\lambda}{6A_\alpha}\left( A_{\alpha} - \int_{\Omega}\phi_\alpha^2(3-2\phi_\alpha) {\rm d}\Omega \right)^2 \nonumber\\
    &+& \int_{\Omega}\textit{g}\phi_\alpha^2\frac{15}{\epsilon^2}\sum_{\beta \neq \alpha }^N \phi_\beta^2\,{\rm d}\Omega
    + \int_{\Omega}\omega\frac{\epsilon^3}{4}\sum_{\beta \neq \alpha }^N B(\phi_\alpha) B(\phi_\beta) \grad \phi_\alpha \cdot \grad\phi_\beta\,{\rm d}\Omega.
 \label{Eqn:energy-functional}
\end{eqnarray} 

\noindent
Here, $\gamma$ is the membrane surface tension, $\lambda$ controls cell-area conservation, and $g$ and $\omega$ denote the intercellular repulsion and adhesion strengths, respectively.
The surface tension term uses the double-well potential
$G(\phi_\alpha)=30\phi_\alpha^2(1-\phi_\alpha)^2$ \cite{gomez2017computational}.
The area of cell $\alpha$ is denoted by $A_\alpha$.
The last term in Eq.~\eqref{Eqn:energy-functional} uses a regularization function \cite{boschman2023fundamental,elliott2009analysis,dziuk2008eulerian,lowengrub2009phase} 
\begin{equation}
    B(\zeta) = \frac{\zeta}{1+\zeta^3},
    \qquad
    \zeta\in[0,1].
\label{Eqn:adhesion-regularization}
\end{equation}
This function is smooth and bounded over the admissible range of the phase field.
Together with the phase field gradients in Eq.~\eqref{Eqn:energy-functional}, it restricts adhesion to regions where neighboring diffuse interfaces overlap.
With the energy functional defined, the evolution of cell $\alpha$ is governed by
\begin{equation}
    \frac{\partial{\phi_\alpha}}{\partial{t}}
    + \boldsymbol{u}_\alpha\cdot\nabla\phi_\alpha 
    =
    -\Gamma \frac{\delta \mathcal{F}}{\delta\phi_\alpha}.
 \label{Eqn:phase-field}
\end{equation} 
Here, $\Gamma$ is a kinetic coefficient that sets the relaxation rate of the phase field in response to the driving forces in the energy functional.
The variational derivative on the right-hand side of Eq.~\eqref{Eqn:phase-field} is
\begin{eqnarray}
    \frac{\delta\mathcal{F}}{\delta \phi_\alpha} 
    &=& \gamma\left(\frac{60}{\epsilon^2}\phi_{\alpha}(1-\phi_{\alpha})(1-2\phi_{\alpha}) 
    - 2\nabla^2\phi_{\alpha}\right) 
    - 2\lambda\left[\frac{\phi_\alpha(1-\phi_\alpha)}{A_{\alpha}}\left(A_{\alpha} 
    - \int_{\Omega}\phi_\alpha^2(3-2\phi_\alpha) \,\text{d}\Omega\right) \right] \nonumber\\
    &+& g\phi_\alpha\frac{60}{\epsilon^2}\sum_{\beta \neq \alpha }^N\phi_\beta^2 \,\,
    - \omega\frac{\epsilon^3}{2} B(\phi_\alpha)  
    \sum_{\beta \neq \alpha }^N \left(
    B(\phi_\beta) \Delta \phi_\beta 
    + B'(\phi_\beta) \abs{\grad\phi_\beta}^2 \right). 
    \label{Eqn:variational}
\end{eqnarray} 
In Eq.~\eqref{Eqn:phase-field}, $\boldsymbol{u}_\alpha$ denotes the velocity of cell $\alpha$, which advects the phase field and is taken to be spatially uniform within cell $\alpha$.
We decompose it into a polarization-driven component $\boldsymbol{u}_\alpha^{P}$ and an interaction-driven component $\boldsymbol{u}_\alpha^{I}$,
\begin{equation}
    \boldsymbol{u}_\alpha
    =
    \boldsymbol{u}_\alpha^{P}
    +
    \boldsymbol{u}_\alpha^{I}.
    \label{Eqn:cell_velocity_decomposition}
\end{equation}
The polarization velocity represents active cell migration and is written as
\begin{equation}
    \boldsymbol{u}_{\alpha}^{P}
    =
    \boldsymbol{u}_{\alpha}^{\mathrm{chem}}
    +
    \boldsymbol{u}_{\alpha}^{\mathrm{CIL}}
    +
    \boldsymbol{u}_{\alpha}^{\mathrm{rand}},
    \label{Eqn:polarization_velocity}
\end{equation}
where $\boldsymbol{u}_{\alpha}^{\mathrm{chem}}$ is the C3a-mediated chemotactic velocity,
$\boldsymbol{u}_{\alpha}^{\mathrm{CIL}}$ is the contact-inhibition velocity, and
$\boldsymbol{u}_{\alpha}^{\mathrm{rand}}$ is a stochastic velocity that represents a persistent random walk. 
The interaction velocity $\boldsymbol{u}_\alpha^{I}$ accounts for mechanical interactions between neighboring cells, including repulsion and adhesion.
These velocity components are further defined in the following subsections.

%


\subsection{\textbf{Chemotaxis}}\label{sec:chemotaxis}

\noindent
The chemotactic velocity $\boldsymbol{u}_{\alpha}^{\mathrm{chem}}$ in Eq.~\eqref{Eqn:polarization_velocity} represents co-attraction, whereby cells migrate toward chemoattractant gradients generated by the same cell population.
In NCC migration, Complement component 3a (C3a), a chemoattractant secreted by NCCs, mediates this mechanism \cite{carmona2011complement}.
Although C3a provides a weak chemotactic bias \cite{theveneau2010collective,bajanca2019vivo}, it helps maintain cohesion within migrating NCC groups \cite{carmona2011complement,theveneau2013chase}.
We denote the C3a concentration by $c(\boldsymbol{x},t)$ and model its evolution through diffusion, secretion, and decay, while neglecting membrane-mediated consumption:
\begin{equation}
\frac{\partial(\psi c)}{\partial t} =
\nabla\cdot\left(D_c\psi\nabla c\right)
+\delta_\psi b_c
-\psi r_c c .
\label{Eqn:chemoattractant}
\end{equation}
Here, $D_c$, $b_c$, and $r_c$ denote the C3a diffusion coefficient, production rate, and decay rate, respectively.
The extracellular and diffuse-membrane markers are defined as
\begin{equation}
\psi = 1-\sum_{\alpha=1}^{N}\phi_\alpha,
\qquad
\delta_\psi = \frac{30}{\epsilon}\psi^2(1-\psi)^2 .
\label{Eqn:extracellular_membrane_markers}
\end{equation}
The marker $\psi$ restricts the transient, diffusion, and decay terms in Eq.~\eqref{Eqn:chemoattractant} to the extracellular space, whereas $\delta_\psi$ localizes C3a secretion to the diffuse cell membrane.
As $\epsilon\to0$, this diffuse-interface formulation recovers the corresponding sharp-interface transport problem with consistent interface conditions \cite{li2009solving}.
Each cell converts the resulting extracellular C3a field into directed motion by sensing the local C3a gradient along its membrane. 
For cell $\alpha$, we compute the membrane-averaged local C3a gradient as
\begin{equation}
\overline{\nabla c}_{\alpha}
=
\frac{\displaystyle\int_{\Omega_\alpha}\delta_{\phi_\alpha}\nabla c\,{\rm d}\Omega}
{\displaystyle\int_{\Omega_\alpha}\delta_{\phi_\alpha}\,{\rm d}\Omega}.
\label{Eqn:membrane_averaged_c3a_gradient}
\end{equation}
Here, $\Omega_\alpha$ is a circular computational window of radius $R_\alpha+1.5\epsilon$ centered at the centroid of cell $\alpha$, where $R_\alpha$ is the equivalent cell radius, see Fig.~\ref{fig:method_schematic}A.
The cell-specific membrane marker is defined as
\begin{equation}
\delta_{\phi_\alpha}
=
\frac{30}{\epsilon}\phi_\alpha^2(1-\phi_\alpha)^2 ,
\label{Eqn:cell_membrane_marker}
\end{equation}
which localizes the average chemotactic gradient in Eq.~\eqref{Eqn:membrane_averaged_c3a_gradient} to the diffuse membrane of cell $\alpha$.
We assume chemotactic velocity to point along $\overline{\nabla c}_{\alpha}$, while its magnitude increases with the sensed gradient strength and remains bounded by the maximum chemotactic speed $v_{\mathrm{chem}}^{\max}$ \cite{colombi2020modelling}:
\begin{equation}
\boldsymbol{u}_{\alpha}^{\mathrm{chem}}
=
\min\left(
\chi\left\lvert\overline{\nabla c}_{\alpha}\right\rvert,
v_{\mathrm{chem}}^{\max}
\right)
\frac{\overline{\nabla c}_{\alpha}}
{\left\lvert\overline{\nabla c}_{\alpha}\right\rvert+\epsilon_{\mathrm{chem}}}.
\label{Eqn:chem_velocity}
\end{equation}
Here, $\chi$ controls sensitivity to the local C3a gradient, and
$\epsilon_{\mathrm{chem}}$ regularizes the normalization when the sensed gradient is small.


\subsection{\textbf{Contact inhibition of locomotion}}
\label{sec:contact-inhibition-of-locomotion}

\noindent
The CIL velocity $\boldsymbol{u}_{\alpha}^{\mathrm{CIL}}$ in Eq.~\eqref{Eqn:polarization_velocity} represents the contribution of contact-mediated repolarization to the polarization velocity of cell $\alpha$.
When a NCC contacts a neighboring cell, protrusions at the contact site retract and new protrusions form away from contact.
This response involves the redistribution of Rho-family GTPases, including Rac1 and RhoA, which promote protrusion and retraction, respectively \cite{roycroft2016molecular,merchant2018rho}.
Rather than resolving Rac1 and RhoA dynamics separately, we introduce a phenomenological scalar field $a(\boldsymbol{x},t)$ that represents their net contact-dependent effect on protrusion-retraction dynamics \cite{raftopoulou2004cell,zegers2014rho}.
Because this signaling response is spatially organized near the cell membrane, we localize $a$ using the diffuse-membrane marker $\delta_\psi$.
The field $a$ represents an effective contact activity signal rather than the concentration of a specific signaling molecule.
\\

We identify cell-cell contact using the overlap measure
\begin{equation}
\Sigma_{\nabla}^{\mathrm{cell}}
=
\sum_{\alpha=1}^{N}
\sum_{\substack{\beta=1\\ \beta\neq\alpha}}^{N}
\left\lvert\nabla\phi_\alpha\right\rvert
\left\lvert\nabla\phi_\beta\right\rvert .
\label{Eqn:cell_contact_measure}
\end{equation}
For simulations with spatial confinement, let $\phi_{\mathrm{bnd}}$ denote the phase field whose diffuse interface represents the confining boundary.
We define the boundary-contact and total contact measures as
\begin{align}
\Sigma_{\nabla}^{\mathrm{bnd}}
&=
\sum_{\alpha=1}^{N}
\left\lvert\nabla\phi_\alpha\right\rvert
\left\lvert\nabla\phi_{\mathrm{bnd}}\right\rvert,
\label{Eqn:boundary_contact_measure}
\\
\Sigma_{\nabla}
&=
\Sigma_{\nabla}^{\mathrm{cell}}
+
\Sigma_{\nabla}^{\mathrm{bnd}}.
\label{Eqn:contact_measure}
\end{align}
Thus, $\Sigma_{\nabla}$ is non-zero wherever diffuse cell-cell or cell-boundary interfaces overlap.
\\

With $a(\boldsymbol{x},0)=0$, the shared contact activity field evolves according to
\begin{equation}
\frac{\partial(\delta_\psi a)}{\partial t}
=
\nabla\cdot\left(D_a\delta_\psi\nabla a\right)
+
\delta_\psi b_a\Sigma_{\nabla}
-
\delta_\psi r_a a
\left(
\frac{\epsilon_a}{\epsilon_a+\Sigma_{\nabla}}
\right).
\label{Eqn:contact_activity}
\end{equation}
Here, $D_a$, $b_a$, and $r_a$ denote the membrane-associated diffusion, production, and decay rates of the contact activity field, respectively.
The production term generates $a$ at cell-cell and cell-boundary contact interfaces.
The factor $\epsilon_a/(\epsilon_a+\Sigma_{\nabla})$ suppresses decay while contact persists and approaches unity after separation, with $\epsilon_a$ as a small value regularizing this transition.
Consequently, $a$ accumulates near persistent contact regions, redistributes slowly along the membrane because $D_a$ is small, and decays after contact is lost.
\\

We use the membrane-localized contact activity field to construct a contact-induced polarity contribution by weighting the local interface normals over the selected contact region.
The polarity contribution to cell $\alpha$ generated by contact with cell $\beta$ is \cite{moure2018three}
\begin{equation}
\boldsymbol{P}_{\alpha\beta}
=
\int_{\Omega_\alpha}
a\,\delta_{\phi_\beta}\,
\nabla\cdot
\left(
\nabla\phi_\alpha\otimes\nabla\phi_\alpha
\right)
\,\mathrm{d}\Omega .
\label{Eqn:pairwise_contact_polarity}
\end{equation}
%
%
Multiplication by $\delta_{\phi_\beta}$ selects the contact activity field associated with the $\alpha$-$\beta$ contact interface, while
$\nabla\cdot\left(\nabla\phi_\alpha\otimes\nabla\phi_\alpha\right)$
provides the local interface-normal direction of cell $\alpha$.
Under the phase field convention used here, $\boldsymbol{P}_{\alpha\beta}$ reorients cell $\alpha$ away from its contact with cell $\beta$.
The net polarity contribution on cell $\alpha$ generated by cell-cell contact is
\begin{equation}
\boldsymbol{P}_{\alpha}^{\mathrm{cell}}
=
\sum_{\beta\in\mathcal{N}(\alpha)}
\boldsymbol{P}_{\alpha\beta},
\label{Eqn:cell_contact_polarity}
\end{equation}
where $\mathcal{N}(\alpha)$ denotes the set of cells neighboring cell $\alpha$.
\\

We model contact with the confining boundary using the same contact activity field and polarity construction, thereby representing a phenomenological CIL-like boundary response \cite{woods2014directional,merchant2018rho,merchant2020rho}.
Replacing the neighboring-cell membrane marker in Eq.~\eqref{Eqn:pairwise_contact_polarity} with the diffuse-boundary marker gives
\begin{equation}
\boldsymbol{P}_{\alpha}^{\mathrm{bnd}}
=
\int_{\Omega_\alpha}
a\,\delta_{\phi_{\mathrm{bnd}}}\,
\nabla\cdot
\left(
\nabla\phi_\alpha\otimes\nabla\phi_\alpha
\right)
\,\mathrm{d}\Omega ,
\label{Eqn:boundary_contact_polarity}
\end{equation}
where
$\delta_{\phi_{\mathrm{bnd}}}
=
\frac{30}{\epsilon}
\phi_{\mathrm{bnd}}^2
(1-\phi_{\mathrm{bnd}})^2$
marks the diffuse interface of the confining boundary.
We use the same parameters that govern contact activity and the polarity response for cell-cell and cell-boundary contacts, treating both as activations of a common effective cell-intrinsic repolarization mechanism.
This parsimonious assumption represents a shared downstream effect on polarity and does not imply identical upstream molecular signaling.
\\

Following previous models of CIL and collective guidance \cite{colombi2020modelling,camley2016collective}, we use a relaxation law so that the CIL velocity responds to contact over a finite time rather than changing instantaneously. 
The total CIL velocity of cell $\alpha$ evolves according to 
\begin{equation}
\frac{\mathrm{d}\boldsymbol{u}_{\alpha}^{\mathrm{CIL}}}{\mathrm{d}t}
=
\frac{1}{\tau_{\text{\tiny CIL}}}
\left[
-\boldsymbol{u}_{\alpha}^{\mathrm{CIL}}
+
\mu
\left(
\boldsymbol{P}_{\alpha}^{\mathrm{cell}}
+
\boldsymbol{P}_{\alpha}^{\mathrm{bnd}}
\right)
\right].
\label{Eqn:cil_velocity}
\end{equation}
Here, $\mu$ sets the strength with which the contact-induced polarity contributions affect the CIL velocity, and $\tau_{\mathrm{CIL}}$ sets the response time.
For a fixed contact configuration, $\boldsymbol{u}_{\alpha}^{\mathrm{CIL}}$ approaches $\mu \left( \boldsymbol{P}_{\alpha}^{\mathrm{cell}} + \boldsymbol{P}_{\alpha}^{\mathrm{bnd}} \right)$, with larger values of $\tau_{\mathrm{CIL}}$ producing a slower response. 
Mechanical adhesion and repulsion enter separately through the interaction velocity $\boldsymbol{u}_{\alpha}^{I}$. 
\\


\subsection{\textbf{Persistent stochastic velocity}}\label{sec:stochastic-velocity}

\noindent
The stochastic component $\boldsymbol{u}_{\alpha}^{\mathrm{rand}}$ in Eq.~\eqref{Eqn:polarization_velocity} accounts for the persistent random motion of NCCs \cite{li2008persistent,bosgraaf2009ordered}.
We represent this behavior using a sequence of cell-specific persistence intervals of length $\Delta T_\alpha^i$.
For cell $\alpha$, the stochastic velocity remains fixed during interval $i$, denoted by
\begin{equation}
\boldsymbol{u}_{\alpha}^{\mathrm{rand},i}
=
\left\lvert\boldsymbol{u}_{\alpha}^{\mathrm{rand},i}\right\rvert
\widehat{\boldsymbol{e}}_{\alpha}^{\mathrm{rand},i},
\qquad
\left\lvert\widehat{\boldsymbol{e}}_{\alpha}^{\mathrm{rand},i}\right\rvert=1.
\label{Eqn:stochastic_velocity}
\end{equation}
Here, $\left\lvert\boldsymbol{u}_{\alpha}^{\mathrm{rand},i}\right\rvert$ is the stochastic speed, and $\widehat{\boldsymbol{e}}_{\alpha}^{\mathrm{rand},i}$ is the unit direction that makes an angle $\theta_{\alpha}^{\mathrm{rand},i}$ with the $x$-axis.
At the end of each persistence interval, we sample the stochastic speed and duration of the next interval as
\begin{equation}
\left\lvert\boldsymbol{u}_{\alpha}^{\mathrm{rand},i+1}\right\rvert
\sim
\mathcal{U}\left(
0.8\overline{u^{\mathrm{rand}}},
1.2\overline{u^{\mathrm{rand}}}
\right),
\qquad
\Delta T_\alpha^{i+1}
\sim
\mathcal{U}\left(
0.8\overline{\Delta T},
1.2\overline{\Delta T}
\right).
\label{Eqn:stochastic_speed_persistence}
\end{equation}
Here, $\mathcal{U}(\ell,r)$ denotes the uniform distribution on $[\ell,r]$, while $\overline{u^{\mathrm{rand}}}$ and $\overline{\Delta T}$ denote the mean stochastic speed and persistence time, respectively, based on experimental measurements \cite{moure2021phase,scarpa2013novel,theveneau2010collective}.
We update the stochastic direction in two steps.
First, we generate an intermediate angle by adding a random turning angle,
\begin{equation}
\theta_{\alpha}^{\mathrm{rand},i+\frac{1}{2}}
=
\theta_{\alpha}^{\mathrm{rand},i}
+
\Delta\theta_{\alpha}^{\mathrm{rand},i+1},
\label{Eqn:random_turning_angle}
\end{equation}
where $\Delta\theta_{\alpha}^{\mathrm{rand},i+1}$ is sampled from the turning-angle distribution described in Fig.~12C of Ref.~\cite{moure2021phase}.
We then bias this intermediate angle toward the current polarization direction to obtain
\begin{equation}
\theta_{\alpha}^{\mathrm{rand},i+1}
=
\theta_{\alpha}^{\mathrm{rand},i+\frac{1}{2}}
+
\rho_{\mathrm{pol}}
\left(
\theta_{\alpha}^{P,i}
-
\theta_{\alpha}^{\mathrm{rand},i+\frac{1}{2}}
\right).
\label{Eqn:stochastic_direction_update}
\end{equation}
Here, $\theta_{\alpha}^{P,i}$ denotes the direction of the polarization velocity $\boldsymbol{u}_\alpha^{P}$ immediately before the update, and $0\leq\rho_{\mathrm{pol}}\leq1$ controls the strength of the directional bias.
%
%
For $\rho_{\mathrm{pol}}=0$, the new direction follows the random turning update, whereas increasing $\rho_{\mathrm{pol}}$ strengthens its bias toward the current polarization direction.
The updated stochastic velocity remains fixed throughout the next persistence interval.


\subsection{\textbf{Interaction velocity}}\label{sec:interaction-velocity}

\noindent
The interaction velocity $\boldsymbol{u}_\alpha^{I}$ in Eq.~\eqref{Eqn:cell_velocity_decomposition} accounts for short-range repulsion and adhesion between cells.
We compute this velocity from the total free-energy functional $\mathcal{F}$ (see Eq.~\eqref{Eqn:total-energy-functional}) as
\begin{equation}
\boldsymbol{u}_\alpha^{I}
=
\xi^{-1}
\int_{\Omega}
\frac{\delta\mathcal{F}}{\delta\phi_\alpha}
\nabla\phi_\alpha\,{\rm d}\Omega,
\label{Eqn:uint_def}
\end{equation}
where $\xi$ is the effective friction coefficient.
Substituting Eq.~\eqref{Eqn:variational} into Eq.~\eqref{Eqn:uint_def}, the surface tension and area conservation contributions reduce to boundary terms and vanish under the imposed phase-field boundary conditions \cite{palmieri2015multiple,wise2008three,moure2021phase}.
The remaining repulsion and adhesion contributions give
\begin{align}
\boldsymbol{u}_\alpha^{I}
&=
\frac{60g}{\xi\epsilon^2}
\int_{\Omega}
\phi_\alpha\nabla\phi_\alpha
\sum_{\substack{\beta=1\\ \beta\neq\alpha}}^{N}
\phi_\beta^2\,{\rm d}\Omega
\nonumber\\
&\quad
-\frac{\omega\epsilon^3}{2\xi}
\int_{\Omega}
B(\phi_\alpha)\nabla\phi_\alpha
\sum_{\substack{\beta=1\\ \beta\neq\alpha}}^{N}
\left[
B(\phi_\beta)\Delta\phi_\beta
+
B'(\phi_\beta)\lvert\nabla\phi_\beta\rvert^2
\right]\,{\rm d}\Omega .
\label{Eqn:uint_expanded}
\end{align}
The first term produces repulsion by penalizing cell-cell overlap, whereas the second term represents adhesion between neighboring diffuse interfaces.





\subsection{\textbf{Numerical implementation}}\label{sec:numerics}

\noindent
We implement the model using the dynamic cluster field (DCF) framework \cite{paspunurwar2024dynamic}.
Although the mathematical model is written in terms of individual phase fields $\phi_\alpha$, DCF reduces computational cost by grouping sufficiently separated cells into cluster fields.
We solve each cluster field on a computational domain defined as the disjoint union of the computational windows $\Omega_\alpha$ (see Fig.~\ref{fig:method_schematic}A) of its constituent cells \cite{paspunurwar2024dynamic}.
The computational window moves with the centroid of cell $\alpha$ and contains its diffuse interface.
Additionally, DCF dynamically reallocates cells among cluster fields so that cells approaching contact are represented by distinct fields, thereby preventing artificial merging within a cluster field.
This construction preserves the underlying single-cell phase-field dynamics while reducing the number of evolved PDE fields from $N$ to $N_c$, where $N_c$ denotes the number of clusters \cite{paspunurwar2024dynamic}.
\\

At each time step, we recover the individual phase fields on their respective computational windows and assemble the derived quantities required by the model, including $\psi$, $\delta_\psi$, $\delta_{\phi_\alpha}$, and $\Sigma_{\nabla}$.
We then evaluate cell-wise quantities, including repulsion and adhesion integrals, contact-mediated polarization response, area-conservation terms, and chemotactic gradients, locally over $\Omega_\alpha$.
In addition to the $N_c$ cluster fields, we solve for the C3a concentration $c$ and contact activity field $a$ throughout $\Omega$.
The coefficients and source terms in these global equations are assembled from the DCF representation of the cell population. 
\\

\noindent
We discretize space using isogeometric analysis (IGA) \cite{hughes2005isogeometric} with spline basis functions provided by PetIGA \cite{dalcin2016petiga}.
We use quadratic $C^1$ B-splines on a uniform Cartesian grid with spacing
$\Delta x=\Delta y=\SI{0.5}{\micro\meter}$ in all simulations.
We set the diffuse-interface width to $\epsilon=\SI{4}{\micro\meter}$, such that approximately eight grid elements resolve each cell interface.
The cluster-field, C3a, and contact activity field equations use the same spline space.
Boundary conditions vary among numerical experiments and are specified with each simulation setup.
The phase field weak form follows our previous DCF formulation \cite{paspunurwar2024dynamic}, while the C3a and contact activity equations use the corresponding Galerkin weak forms.
\\

\noindent
Because $\psi$ weights the transport operators in Eq.~\eqref{Eqn:chemoattractant}, values approaching zero inside cells can produce ill-conditioned discrete operators.
We therefore replace $\psi$ in the numerical formulation by the smooth positive marker
\begin{equation}
\psi_{\mathrm{sm}}
=
\frac{1}{2}
\left[
\psi+\psi_{\min}
+\sqrt{(\psi-\psi_{\min})^2+\eta_\psi}
\right],
\qquad
\psi_{\min}=10^{-8},
\quad
\eta_\psi=4\times10^{-12}.
\label{Eqn:smooth_extracellular_marker}
\end{equation}
This expression provides a smooth approximation of
$\max(\psi,\psi_{\min})$ \cite{biswas2021smu},
where $\psi_{\min}$ is the nominal lower bound and $\eta_\psi$ controls the smoothing.
For the chosen parameters, $\psi_{\mathrm{sm}}\approx10^{-6}$ inside cells, where $\psi\approx0$, and $\psi_{\mathrm{sm}}\approx1$ in the extracellular space away from the diffuse interfaces, where $\psi\approx1$.
Subsequently, we compute the diffuse membrane marker from the regularized field:
\begin{equation}
\delta_\psi
=
\frac{30}{\epsilon}
\psi_{\mathrm{sm}}^2(1-\psi_{\mathrm{sm}})^2.
\label{Eqn:smooth_membrane_marker}
\end{equation}

\noindent
We advance the cluster field, C3a, and contact activity equations as a coupled system using the forward Euler method with a fixed time step
$\Delta t=\SI{0.0025}{\second}$.
At time level $t^n$, all coupling terms in these equations are evaluated from the available field solutions and held fixed while each equation is advanced to $t^{n+1}$.
The updated cluster fields, C3a concentration, and contact activity field are then used to recompute the coupling terms at the beginning of the next time step.
Thus, coupling among the three field equations remains fully explicit.
To reduce memory requirements and avoid global mass-matrix solves, we replace the consistent mass matrices with lumped diagonal approximations \cite{duczek2019mass,gravenkamp2020mass}.
Mass-matrix lumping allows inexpensive inversion and enables efficient simulation of large multicellular systems \cite{paspunurwar2024dynamic}.
The selected time step satisfies the stability restrictions associated with the explicit scheme and diffuse-interface dynamics.
%
\\

\noindent
Cell-wise velocity components are also evaluated explicitly at each time level.
The chemotactic velocity $\boldsymbol{u}_{\alpha}^{\mathrm{chem}}$ is recomputed from the membrane-averaged C3a gradient, and the CIL velocity $\boldsymbol{u}_{\alpha}^{\mathrm{CIL}}$ is advanced using Eq.~\eqref{Eqn:cil_velocity}.
The stochastic velocity $\boldsymbol{u}_{\alpha}^{\mathrm{rand}}$ remains fixed over each persistence interval and is updated when that interval ends.
We then assemble the polarization velocity using Eq.~\eqref{Eqn:polarization_velocity} and the total cell velocity using Eq.~\eqref{Eqn:cell_velocity_decomposition}.
\\

\noindent
Together, DCF grouping, IGA discretization, mass-matrix lumping, and explicit time integration enable efficient multicellular simulations while retaining the underlying single-cell phase-field dynamics \cite{paspunurwar2024dynamic,paspunurwar2025decoding}.


\subsection{\textbf{Experimental pairwise collision dynamics constrain model parameters}} \label{sec:parametric-study} \noindent

\begin{figure}[h!] 

\centering 
\includegraphics[width=\linewidth]{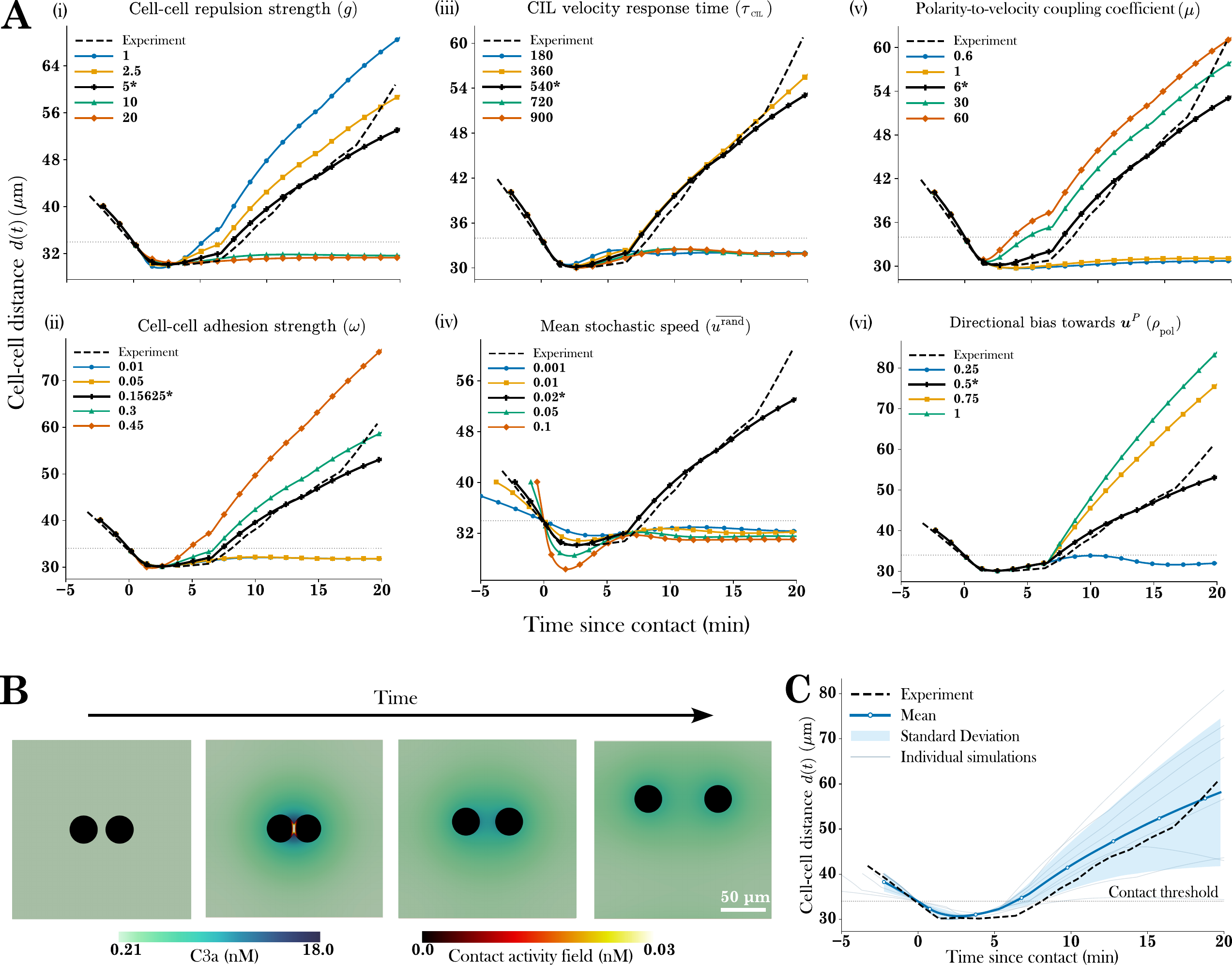}
\caption{ 
\textbf{Independent parameter variation constrains pairwise collision dynamics.}
\textbf{(A)} Deterministic cell center-to-center distance, $d(t)$, during independent variations in {(i)} cell-cell repulsion strength $g$, {(ii)} cell-cell adhesion strength $\omega$, {(iii)} CIL response time $\tau_{\text{\tiny CIL}}$, {(iv)} mean stochastic speed $\overline{u^{\mathrm{rand}}}$, {(v)} polarity-to-velocity coupling coefficient $\mu$, and {(vi)} directional bias $\rho_{\mathrm{pol}}$. 
%
%
The dashed black curve shows the experimental trajectory extracted from Ref.~\cite{scarpa2013novel}.
%
%
\textbf{(B)} Representative snapshots of cell approach, contact, and separation using the selected parameter set, together with the evolving C3a concentration and contact activity field. 
\textbf{(C)} Pairwise collision trajectories from eight simulations after restoring stochastic motility. 
Thin blue curves show individual simulations, the solid blue curve shows the mean across the eight simulations, and the shaded region denotes one standard deviation.
The dashed black curve shows the experimental reference \cite{scarpa2013novel}, and the horizontal line marks the contact threshold.
} 
\label{fig:validation} 
\end{figure}

Pairwise collision experiments provide a direct basis for constraining the parameters that govern cell contact, repolarization, and separation after contact.
We adopt the remaining model parameters from the literature, as listed in Table~\ref{Table:params}, and examine the cell-cell repulsion strength $g$, cell-cell adhesion strength $\omega$, CIL response time $\tau_{\mathrm{CIL}}$, mean stochastic speed $\overline{u^{\mathrm{rand}}}$, polarity-to-velocity coupling coefficient $\mu$, and directional bias $\rho_{\mathrm{pol}}$.
We first perform deterministic simulations by varying each parameter independently while holding all others at their asterisked values; see Fig.~\ref{fig:validation}A, where the black curve in each panel corresponds to the asterisked value of the varied parameter, while the colored curves correspond to different values.
These parameter variations identify how each parameter affects the approach, contact, and subsequent separation phases of the collision.
Guided by these trends, we restore stochastic motility and select the parameter combination whose mean trajectory across simulations most closely reproduces the experimentally measured cell-cell distance reported by Scarpa et al.~\cite{scarpa2013novel}.
\\

\noindent
We initialize two NCCs symmetrically in a $\SI{200}{\micro\meter}\times\SI{200}{\micro\meter}$ domain with an initial center-to-center distance of $\SI{40}{\micro\meter}$.
The cells have equal and opposite initial velocities directed toward one another, each with magnitude $\SI{0.020}{\micro\meter\per\second}$.
We initialize the chemical fields as $c(\boldsymbol{x},0)=0$ and $a(\boldsymbol{x},0)=0$ and simulate $\SI{40}{\minute}$ of migration with self-generated chemotaxis and cell-cell CIL active.
We measure the center-to-center distance $d(t)$ and align each trajectory such that $t=0$ corresponds to first contact, defined by $d\leq\SI{34}{\micro\meter}$ following Ref.~\cite{scarpa2013novel}.
We compare the approach trajectory, minimum separation, contact duration, and subsequent separation with the experimental curve.
\\

We set $\lvert\boldsymbol{u}_{\alpha}^{\mathrm{rand}}\rvert=\overline{u^{\mathrm{rand}}}$, $\Delta T_\alpha=\overline{\Delta T}$, and $\Delta\theta_{\alpha}^{\mathrm{rand}}=0$ throughout these deterministic parameter variations.
After selecting the final parameter combination, we restore the stochastic motility updates and perform eight independent simulations to evaluate the resulting distribution of pairwise collision trajectories.

\paragraph{Cell-cell repulsion strength $g$}

\noindent
The repulsion strength controls the overlap between diffuse cell interfaces and therefore indirectly regulates the production of contact activity.
Reducing $g$ permits greater interfacial overlap, increases $\Sigma_{\nabla}^{\mathrm{cell}}$, and causes separation to begin earlier and proceed more rapidly than observed experimentally (Fig.~\ref{fig:validation}A(i)).
Increasing $g$ reduces overlap and weakens contact activity production, causing the cells to remain near the contact threshold rather than separate.

\paragraph{Cell-cell adhesion strength $\omega$}

\noindent
Adhesion opposes cell-cell repulsion and increases the extent and duration of interfacial overlap.
Small values of $\omega$ generate insufficient contact activity to sustain separation, whereas large values increase the accumulation of $a$ and produce excessively rapid separation (Fig.~\ref{fig:validation}A(ii)).

\paragraph{CIL response time $\tau_{\mathrm{CIL}}$}

\noindent
The response time controls how rapidly the CIL velocity adjusts to changes in the contact-induced polarity input.
Small values cause the CIL velocity to respond and decay rapidly as contact changes, producing a brief response that does not sustain separation.
Large values slow the response to contact and delay the reversal of cell motion (Fig.~\ref{fig:validation}A(iii)).
Intermediate values balance responsiveness and persistence.

\paragraph{Mean stochastic speed $\overline{u^{\mathrm{rand}}}$}

\noindent
The mean stochastic speed controls the approach rate and the inward motility that the CIL response must overcome after contact.
At low speeds, the cells develop insufficient interfacial overlap and contact activity to sustain separation after contact.
At high speeds, the cells approach and overlap more rapidly, but the stronger inward motility opposes reversal and prolongs contact (Fig.~\ref{fig:validation}A(iv)).

\paragraph{Polarity-to-velocity coupling coefficient $\mu$}

\noindent
The coefficient $\mu$ scales the contact-induced polarity input to the CIL velocity in Eq.~\eqref{Eqn:cil_velocity}.
Since contact activity scales with $b_a$, the CIL velocity input depends on the product $b_a\mu$, and collision trajectories cannot determine these parameters independently.
We therefore fix $b_a$ and calibrate $\mu$ using the pairwise collision data.
Small values produce an insufficient response and leave the cells near the contact threshold, whereas increasing $\mu$ strengthens the reversal and increases the rate of separation after contact (Fig.~\ref{fig:validation}A(v)).

\paragraph{Directional bias $\rho_{\mathrm{pol}}$}

\noindent
The parameter $\rho_{\mathrm{pol}}$ controls how strongly renewed stochastic polarization follows the current polarization velocity.
It has little effect during approach because both cells initially move toward one another.
After CIL reorients the cells, larger values of $\rho_{\mathrm{pol}}$ preserve the outward polarity and increase the separation rate, whereas small values do not sustain the reversal (Fig.~\ref{fig:validation}A(vi)).
Excessively large values yield greater persistence after contact than is observed experimentally.
\\

\noindent
Together, these variations show that the collision trajectory is governed by complementary mechanical, signaling, and motility processes.
The parameters $g$ and $\omega$ regulate interfacial overlap and contact activity production, $\tau_{\mathrm{CIL}}$ determines the response timing, $\overline{u^{\mathrm{rand}}}$ controls the collision kinematics, $\mu$ sets the strength of contact-induced repolarization, and $\rho_{\mathrm{pol}}$ governs persistence after reversal.
Guided by these trends, we select a baseline parameter combination that jointly reproduces the experimental approach, minimum separation, contact duration, and subsequent separation in the deterministic comparison, rather than optimizing each parameter independently.
When stochastic motility is restored, the eight simulations retain the experimentally observed collision sequence while producing variation in the separation rate after contact (Fig.~\ref{fig:validation}B and C).
The mean trajectory across the simulations closely follows the experimental trajectory, which remains within the predicted variation among simulations.
The selected parameter values used for the stochastic simulations in Fig.~\ref{fig:validation}C are listed in Table~\ref{Table:params} and are retained in all subsequent simulations.
\\


\begin{table}[!ht]
\centering
\caption{Model and numerical parameters used in subsequent simulations.}
\label{Table:params}
\small
\renewcommand{\arraystretch}{1.15}

\begin{tabularx}{\textwidth}{
@{}
l
>{\hsize=1.4\hsize\raggedright\arraybackslash}X
l
>{\hsize=0.6\hsize\raggedright\arraybackslash}X
@{}
}\toprule
\textbf{Symbol}
& \textbf{Description}
& \textbf{Value}
& \textbf{Reference} \\
\midrule

\multicolumn{4}{@{}l}{\textit{Cell geometry and phase-field dynamics}} \\

$R$
& Equivalent cell radius
& \SI{15}{\micro\meter}
& \cite{theveneau2013chase,colombi2020modelling} \\

$\Gamma$
& Phase-field relaxation coefficient
& \SI{0.38}{\micro\meter^{2}\,\pico\newton^{-1}\,\second^{-1}}
& \cite{kulawiak2016modeling} \\

$\gamma$
& Membrane surface tension coefficient
& \SI{3.6}{\pico\newton}
& \cite{kulawiak2016modeling} \\

$\lambda$
& Cell-area conservation strength
& \SI{100}{\pico\newton\,\micro\meter^{-2}}
& \cite{moure2021phase,fuller2010external} \\

$g$
& Cell-cell repulsion strength
& \SI{5}{\pico\newton}
& Calibrated  \\

$\omega$
& Cell-cell adhesion strength
& \SI{0.3}{\pico\newton\,\micro\meter^{-3}}
& Calibrated  \\

$\xi$
& Effective friction coefficient
& \SI{100.0}{\pico\newton\,\second\,\micro\meter^{-2}}
& \cite{kulawiak2016modeling} \\

$\epsilon$
& Diffuse-interface width
& \SI{4}{\micro\meter}
& \cite{paspunurwar2024dynamic,moure2021phase} \\

\midrule
\multicolumn{4}{@{}l}{\textit{Chemotaxis}} \\

$D_c$
& C3a diffusion coefficient
& \SI{8}{\micro\meter^{2}\,\second^{-1}}
& \cite{szabo2016vivo,woods2014directional} \\

$b_c$
& C3a production rate
& \SI{2.7746}{\nano\Molar\,\micro\meter\,\second^{-1}}
& \cite{szabo2016vivo,woods2014directional} \\

$r_c$
& C3a decay rate
& \SI{0.006}{\second^{-1}}
& \cite{szabo2016vivo,woods2014directional} \\

$\chi$
& Chemotactic sensitivity
& \SI{0.01}{\micro\meter^{2}\,\nano\Molar^{-1}\,\second^{-1}}
& Estimated\textsuperscript{a} \\

$v_{\mathrm{chem}}^{\max}$
& Maximum chemotactic speed
& \SI{0.0463}{\micro\meter\,\second^{-1}}
& \cite{colombi2020modelling} \\

$\epsilon_{\mathrm{chem}}$ 
& Regularization factor for chemotactic velocity 
& \SI{1e-8}{\nano\Molar\,\micro\meter^{-1}}
& Estimated\\ 

\midrule
\multicolumn{4}{@{}l}{\textit{Contact inhibition of locomotion}} \\

$D_a$
& Contact activity diffusion coefficient
& \SI{0.1}{\micro\meter^{2}\,\second^{-1}}
& \cite{das2015single} \\

$b_a$
& Contact activity production coefficient
& \SI{0.008}{\nano\Molar\,\micro\meter^{2}\,\second^{-1}}
& Estimated\textsuperscript{b} \\

$r_a$
& Contact activity decay rate
& \SI{0.5}{\second^{-1}}
& Estimated\textsuperscript{c} \\

$\mu$
& Polarity-to-velocity coupling coefficient
& \SI{6}{\micro\meter^{3}\,\nano\Molar^{-1}\,\second^{-1}}
& Calibrated \\

$\tau_{\text{\tiny CIL}}$
& CIL velocity response time
& \SI{540}{\second}
& Calibrated \\

$\epsilon_a$ 
& Regularization factor for contact activity field evolution dynamics 
& \SI{1e-8}{\micro\meter^{-2}}
& Estimated\\ 

\midrule
\multicolumn{4}{@{}l}{\textit{Persistent stochastic motion}} \\

$\overline{u^{\mathrm{rand}}}$
& Mean stochastic speed
& \SI{0.02}{\micro\meter\,\second^{-1}}
& Calibrated \\

$\overline{\Delta T}$
& Mean persistence interval
& \begin{tabular}[t]{@{}l@{}}
    \SI{450}{\second} (single-cell)\\
    \SI{900}{\second} (collective migration)
  \end{tabular}
& \cite{theveneau2010collective} \\

$\rho_{\mathrm{pol}}$
& Directional bias toward polarization velocity
& $0.5$
& Calibrated \\

\midrule
\multicolumn{4}{@{}l}{\textit{Numerical parameters}} \\

$\Delta x=\Delta y$
& Element size
& \SI{0.5}{\micro\meter}
& Diffuse-interface resolution \\

$\Delta t$
& Time-step size
& \SI{0.0025}{\second}
& Forward Euler Stability constraint \\

\bottomrule
\multicolumn{4}{@{}p{\textwidth}@{}}{%
\footnotesize\raggedright
\textsuperscript{a}
The collective simulations are weakly sensitive to $\chi$ because large C3a gradients drive the chemotactic speed to $v_{\mathrm{chem}}^{\max}$.
\par
\textsuperscript{b} The value of $b_a$ sets the contact activity scale.
Since the CIL velocity input depends on the product $b_a\mu$, we calibrate $\mu$ while holding $b_a$ fixed rather than fitting both parameters independently.
\par
\textsuperscript{c}
Rapid decay is chosen to prevent residual contact activity field from continuing to drive repolarization after cell-cell separation. With $r_a=\SI{0.5}{\per\second}$, the decay reaction removes approximately $99\%$ of the activity within $\SI{9.2}{\second}$ after contact is lost.
\par
} \\
\end{tabularx}
\end{table}


\clearpage

\subsection{\textbf{Declaration of generative AI usage}}
During the preparation of this manuscript, the authors used ChatGPT 6 Pro in the initial versions to improve grammar, clarity, and readability.
Following its use, the authors carefully reviewed and edited the content to ensure accuracy and take full responsibility for the final version of the manuscript.


\section{Acknowledgements}

This research was partially funded by the National Science Foundation (Award no. CMMI 1952912). The opinions, findings, conclusions, or recommendations expressed in this study are solely those of the authors and do not necessarily represent the views of the NSF.  This work uses the Bridges-2 system at the Pittsburgh Supercomputing Center (PSC) through allocation no. MCH220014 from the Advanced Cyberinfrastructure Coordination Ecosystem: Services and Support (ACCESS) program, which is supported by the National Science Foundation, grant nos. 2325419 and 1952912.



\section{Data Availability}

Data sets generated during the current study will be made available from the corresponding author on reasonable request. 


\section{Competing interests}

Authors declare no competing interests.

\section{Author Contributions}

Conceptualization: A.P. and H.G.; Methodology: A.P. and H.G.; Software: A.P.; Analysis: A.P. and H.G.; Writing - original draft: A.P. and H.G.; Writing - review and editing: A.P. and H.G.

\newpage

\bibliographystyle{elsarticle-num}
\bibliography{reference}

@article{das2015single,
  title={Single-molecule tracking of small GTPase Rac1 uncovers spatial regulation of membrane translocation and mechanism for polarized signaling},
  author={Das, Sulagna and Yin, Taofei and Yang, Qingfen and Zhang, Jingqiao and Wu, Yi I and Yu, Ji},
  journal={Proceedings of the National Academy of Sciences},
  volume={112},
  number={3},
  pages={E267--E276},
  year={2015},
  publisher={National Academy of Sciences}
}

@article{biswas2021smu,
  title={SMU: smooth activation function for deep networks using smoothing maximum technique},
  author={Biswas, Koushik and Kumar, Sandeep and Banerjee, Shilpak and Pandey, Ashish Kumar},
  journal={arXiv preprint arXiv:2111.04682},
  year={2021}
}

@article{labonne1998induction,
  title={Induction and patterning of the neural crest, a stem cell-like precursor population},
  author={LaBonne, Carole and Bronner-Fraser, Marianne},
  journal={Journal of neurobiology},
  volume={36},
  number={2},
  pages={175--189},
  year={1998},
  publisher={Wiley Online Library}
}

@article{leathers2022time,
  title={Time to go: neural crest cell epithelial-to-mesenchymal transition},
  author={Leathers, Tess A and Rogers, Crystal D},
  journal={Development},
  volume={149},
  number={15},
  pages={dev200712},
  year={2022},
  publisher={The Company of Biologists Ltd}
}

@article{mayor1995induction,
  title={Induction of the prospective neural crest of Xenopus},
  author={Mayor, Roberto and Morgan, Richard and Sargent, Michael G},
  journal={Development},
  volume={121},
  number={3},
  pages={767--777},
  year={1995},
  publisher={The Company of Biologists Ltd}
}

@article{li2008persistent,
  title={Persistent cell motion in the absence of external signals: a search strategy for eukaryotic cells},
  author={Li, Liang and N{\o}rrelykke, Simon F and Cox, Edward C},
  journal={PLoS one},
  volume={3},
  number={5},
  pages={e2093},
  year={2008},
  publisher={Public Library of Science San Francisco, USA}
}

@article{mclennan2010vascular,
  title={Vascular endothelial growth factor (VEGF) regulates cranial neural crest migration in vivo},
  author={McLennan, Rebecca and Teddy, Jessica M and Kasemeier-Kulesa, Jennifer C and Romine, Morgan H and Kulesa, Paul M},
  journal={Developmental biology},
  volume={339},
  number={1},
  pages={114--125},
  year={2010},
  publisher={Elsevier}
}

@article{thiery2009epithelial,
  title={Epithelial-mesenchymal transitions in development and disease},
  author={Thiery, Jean Paul and Acloque, Herv{\'e} and Huang, Ruby YJ and Nieto, M Angela},
  journal={cell},
  volume={139},
  number={5},
  pages={871--890},
  year={2009},
  publisher={Elsevier}
}

@article{bajanca2019vivo,
  title={In vivo topology converts competition for cell-matrix adhesion into directional migration},
  author={Bajanca, Fernanda and Gouignard, Nad{\`e}ge and Colle, Charlotte and Parsons, Maddy and Mayor, Roberto and Theveneau, Eric},
  journal={Nature communications},
  volume={10},
  number={1},
  pages={1518},
  year={2019},
  publisher={Nature Publishing Group UK London}
}

@article{colombi2020modelling,
  title={Modelling chase-and-run migration in heterogeneous populations},
  author={Colombi, Annachiara and Scianna, Marco and Painter, KEVIN JOHN and Preziosi, Luigi},
  journal={Journal of Mathematical Biology},
  volume={80},
  number={1},
  pages={423--456},
  year={2020},
  publisher={Springer}
}

@article{theveneau2013chase,
  title={Chase-and-run between adjacent cell populations promotes directional collective migration},
  author={Theveneau, Eric and Steventon, Benjamin and Scarpa, Elena and Garcia, Simon and Trepat, Xavier and Streit, Andrea and Mayor, Roberto},
  journal={Nature cell biology},
  volume={15},
  number={7},
  pages={763--772},
  year={2013},
  publisher={Nature Publishing Group UK London}
}

@article{theveneau2013collective,
  title={Collective cell migration of epithelial and mesenchymal cells},
  author={Theveneau, Eric and Mayor, Roberto},
  journal={Cellular and Molecular Life Sciences},
  volume={70},
  number={19},
  pages={3481--3492},
  year={2013},
  publisher={Springer}
}

@article{wynn2012computational,
  title={Computational modelling of cell chain migration reveals mechanisms that sustain follow-the-leader behaviour},
  author={Wynn, Michelle L and Kulesa, Paul M and Schnell, Santiago},
  journal={Journal of The Royal Society Interface},
  volume={9},
  number={72},
  pages={1576--1588},
  year={2012},
  publisher={The Royal Society}
}

@article{kabla2012collective,
  title={Collective cell migration: leadership, invasion and segregation},
  author={Kabla, Alexandre J},
  journal={Journal of The Royal Society Interface},
  volume={9},
  number={77},
  pages={3268--3278},
  year={2012},
  publisher={The Royal Society}
}

@article{carmona2011complement,
  title={Complement fragment C3a controls mutual cell attraction during collective cell migration},
  author={Carmona-Fontaine, Carlos and Theveneau, Eric and Tzekou, Apostolia and Tada, Masazumi and Woods, Mae and Page, Karen M and Parsons, Maddy and Lambris, John D and Mayor, Roberto},
  journal={Developmental cell},
  volume={21},
  number={6},
  pages={1026--1037},
  year={2011},
  publisher={Elsevier}
}

@article{carmona2008contact,
  title={Contact inhibition of locomotion in vivo controls neural crest directional migration},
  author={Carmona-Fontaine, Carlos and Matthews, Helen K and Kuriyama, Sei and Moreno, Mauricio and Dunn, Graham A and Parsons, Maddy and Stern, Claudio D and Mayor, Roberto},
  journal={Nature},
  volume={456},
  number={7224},
  pages={957--961},
  year={2008},
  publisher={Nature Publishing Group UK London}
}

@article{camley2016collective,
  title={Collective signal processing in cluster chemotaxis: Roles of adaptation, amplification, and co-attraction in collective guidance},
  author={Camley, Brian A and Zimmermann, Juliane and Levine, Herbert and Rappel, Wouter-Jan},
  journal={PLoS computational biology},
  volume={12},
  number={7},
  pages={e1005008},
  year={2016},
  publisher={Public Library of Science San Francisco, CA USA}
}

@article{camley2017physical,
  title={Physical models of collective cell motility: from cell to tissue},
  author={Camley, Brian A and Rappel, Wouter-Jan},
  journal={Journal of physics D: Applied physics},
  volume={50},
  number={11},
  pages={113002},
  year={2017},
  publisher={IOP Publishing}
}

@article{palmieri2015multiple,
  title={Multiple scale model for cell migration in monolayers: Elastic mismatch between cells enhances motility},
  author={Palmieri, Benoit and Bresler, Yony and Wirtz, Denis and Grant, Martin},
  journal={Scientific reports},
  volume={5},
  number={1},
  pages={11745},
  year={2015},
  publisher={Nature Publishing Group UK London}
}

@article{merchant2018rho,
  title={A Rho-GTPase based model explains spontaneous collective migration of neural crest cell clusters},
  author={Merchant, Brian and Edelstein-Keshet, Leah and Feng, James J},
  journal={Developmental biology},
  volume={444},
  pages={S262--S273},
  year={2018},
  publisher={Elsevier}
}

@article{merchant2020rho,
  title={A Rho-GTPase based model explains group advantage in collective chemotaxis of neural crest cells},
  author={Merchant, Brian and Feng, James J},
  journal={Physical Biology},
  volume={17},
  number={3},
  pages={036002},
  year={2020},
  publisher={IOP Publishing}
}

@article{kulawiak2016modeling,
  title={Modeling contact inhibition of locomotion of colliding cells migrating on micropatterned substrates},
  author={Kulawiak, Dirk Alexander and Camley, Brian A and Rappel, Wouter-Jan},
  journal={PLoS computational biology},
  volume={12},
  number={12},
  pages={e1005239},
  year={2016},
  publisher={Public Library of Science San Francisco, CA USA}
}

@phdthesis{boschman2023fundamental,
  title={Fundamental Theories for Bulk-Surface Interactions: Evolving Interfaces, Phase-Field Adhesion and Mixed-Dimensional Flows},
  author={Boschman, Anne M},
  year={2023},
  school={University of Nottingham (United Kingdom)}
}

@article{sauka2008evolution,
  title={Evolution of the neural crest viewed from a gene regulatory perspective},
  author={Sauka-Spengler, Tatjana and Bronner-Fraser, Marianne},
  journal={genesis},
  volume={46},
  number={11},
  pages={673--682},
  year={2008},
  publisher={Wiley Online Library}
}

@article{abercrombie1953observations,
  title={Observations on the social behaviour of cells in tissue culture: I. Speed of movement of chick heart fibroblasts in relation to their mutual contacts},
  author={Abercrombie, Michael and Heaysman, Joan EM},
  journal={Experimental cell research},
  volume={5},
  number={1},
  pages={111--131},
  year={1953},
  publisher={Elsevier}
}

@article{abercrombie1954observations,
  title={Observations on the social behaviour of cells in tissue culture: II.“Monolayering” of fibroblasts},
  author={Abercrombie, Michael and Heaysman, Joan EM},
  journal={Experimental cell research},
  volume={6},
  number={2},
  pages={293--306},
  year={1954},
  publisher={Elsevier}
}

@article{lowengrub2009phase,
  title={Phase-field modeling of the dynamics of multicomponent vesicles: Spinodal decomposition, coarsening, budding, and fission},
  author={Lowengrub, John S and R{\"a}tz, Andreas and Voigt, Axel},
  journal={Physical Review E—Statistical, Nonlinear, and Soft Matter Physics},
  volume={79},
  number={3},
  pages={031926},
  year={2009},
  publisher={APS}
}

@article{elliott2009analysis,
  title={Analysis of a diffuse interface approach to an advection diffusion equation on a moving surface},
  author={Elliott, Charles M and Stinner, Bj{\"o}rn},
  journal={Mathematical Models and Methods in Applied Sciences},
  volume={19},
  number={05},
  pages={787--802},
  year={2009},
  publisher={World Scientific}
}

@article{dziuk2008eulerian,
  title={Eulerian finite element method for parabolic PDEs on implicit surfaces},
  author={Dziuk, Gerhard and Elliott, Charles M},
  journal={Interfaces and Free Boundaries},
  volume={10},
  number={1},
  pages={119--138},
  year={2008}
}

@article{scarpa2013novel,
  title={A novel method to study contact inhibition of locomotion using micropatterned substrates},
  author={Scarpa, Elena and Roycroft, Alice and Theveneau, Eric and Terriac, Emmanuel and Piel, Matthieu and Mayor, Roberto},
  journal={Biology open},
  volume={2},
  number={9},
  pages={901--906},
  year={2013},
  publisher={The Company of Biologists Bidder Building, 140 Cowley Road, Cambridge, CB4~…}
}

@article{paspunurwar2025decoding,
  title={Decoding complex transport patterns in flow-induced autologous chemotaxis of multicellular systems},
  author={Paspunurwar, Aditya Shankar and Gomez, Hector},
  journal={Biomechanics and Modeling in Mechanobiology},
  volume={24},
  number={1},
  pages={197--212},
  year={2025},
  publisher={Springer}
}

@article{woods2014directional,
  title={Directional collective cell migration emerges as a property of cell interactions},
  author={Woods, Mae L and Carmona-Fontaine, Carlos and Barnes, Chris P and Couzin, Iain D and Mayor, Roberto and Page, Karen M},
  journal={PLoS One},
  volume={9},
  number={9},
  pages={e104969},
  year={2014},
  publisher={Public Library of Science San Francisco, USA}
}

@article{camley2016emergent,
  title={Emergent collective chemotaxis without single-cell gradient sensing},
  author={Camley, Brian and Zimmermann, Juliane and Levine, Herbert and Rappel, Wouter-Jan},
  journal={Biophysical Journal},
  volume={110},
  number={3},
  pages={306a},
  year={2016},
  publisher={Elsevier}
}

@article{trepat2009physical,
  title={Physical forces during collective cell migration},
  author={Trepat, Xavier and Wasserman, Michael R and Angelini, Thomas E and Millet, Emil and Weitz, David A and Butler, James P and Fredberg, Jeffrey J},
  journal={Nature physics},
  volume={5},
  number={6},
  pages={426--430},
  year={2009},
  publisher={Nature Publishing Group UK London}
}

@article{szabo2016vivo,
  title={In vivo confinement promotes collective migration of neural crest cells},
  author={Szab{\'o}, Andr{\'a}s and Melchionda, Manuela and Nastasi, Giancarlo and Woods, Mae L and Campo, Salvatore and Perris, Roberto and Mayor, Roberto},
  journal={Journal of Cell Biology},
  volume={213},
  number={5},
  pages={543--555},
  year={2016},
  publisher={The Rockefeller University Press}
}

@article{szabo2016modelling,
  title={Modelling collective cell migration of neural crest},
  author={Szab{\'o}, Andr{\'a}s and Mayor, Roberto},
  journal={Current opinion in cell biology},
  volume={42},
  pages={22--28},
  year={2016},
  publisher={Elsevier}
}

@article{chen1998fiber,
  title={A fiber matrix model for interstitial fluid flow and permeability in ligaments and tendons},
  author={Chen, Chih-Tung and Malkus, David S and Vanderby Jr, Ray},
  journal={Biorheology},
  volume={35},
  number={2},
  pages={103--118},
  year={1998},
  publisher={Elsevier}
}

@article{ridley2015rho,
  title={Rho GTPase signalling in cell migration},
  author={Ridley, Anne J},
  journal={Current opinion in cell biology},
  volume={36},
  pages={103--112},
  year={2015},
  publisher={Elsevier}
}

@article{roycroft2016molecular,
  title={Molecular basis of contact inhibition of locomotion},
  author={Roycroft, Alice and Mayor, Roberto},
  journal={Cellular and Molecular Life Sciences},
  volume={73},
  number={6},
  pages={1119--1130},
  year={2016},
  publisher={Springer}
}

@article{paspunurwar2024dynamic,
  title={Dynamic cluster field modeling of collective chemotaxis},
  author={Paspunurwar, Aditya Shankar and Moure, Adrian and Gomez, Hector},
  journal={Scientific Reports},
  volume={14},
  number={1},
  pages={25162},
  year={2024},
  publisher={Nature Publishing Group UK London}
}

@article{li2009solving,
  title={Solving PDEs in complex geometries: a diffuse domain approach},
  author={Li, X and Lowengrub, John and R{\"a}tz, A and Voigt, A25128341178},
  journal={Communications in mathematical sciences},
  volume={7},
  number={1},
  pages={81},
  year={2009},
  publisher={NIH Public Access}
}

@article{moure2018three,
  title={Three-dimensional simulation of obstacle-mediated chemotaxis},
  author={Moure, Adrian and Gomez, Hector},
  journal={Biomechanics and modeling in mechanobiology},
  volume={17},
  pages={1243--1268},
  year={2018},
  publisher={Springer}
}

@ARTICLE{moure2021phase,
  title={Phase-field modeling of individual and collective cell migration},
  author={Moure, Adrian and Gomez, Hector},
  journal={Archives of Computational Methods in Engineering},
  volume={28},
  pages={311--344},
  year={2021},
  publisher={Springer}
}

@ARTICLE{anderson1989colloid,
  title={Colloid transport by interfacial forces},
  author={Anderson, John L},
  journal={Annual review of fluid mechanics},
  volume={21},
  number={1},
  pages={61--99},
  year={1989},
  publisher={Annual Reviews 4139 El Camino Way, PO Box 10139, Palo Alto, CA 94303-0139, USA}
}

@article{gomez2019review,
  title={A review on computational modelling of phase-transition problems},
  author={Gomez, Hector and Bures, Miguel and Moure, Adrian},
  journal={Philosophical Transactions of the Royal Society A},
  volume={377},
  number={2143},
  pages={20180203},
  year={2019},
  publisher={The Royal Society Publishing}
}

@article{gomez2017computational,
  title={Computational phase-field modeling},
  author={Gomez, Hector and Van der Zee, Kristoffer G},
  journal={Encyclopedia of computational mechanics},
  pages={1--35},
  year={2017},
  publisher={John Wiley \& Sons, Ltd Hoboken, NJ}
}

@article{theveneau2010collective,
  title={Collective chemotaxis requires contact-dependent cell polarity},
  author={Theveneau, Eric and Marchant, Lorena and Kuriyama, Sei and Gull, Mazhar and Moepps, Barbara and Parsons, Maddy and Mayor, Roberto},
  journal={Developmental cell},
  volume={19},
  number={1},
  pages={39--53},
  year={2010},
  publisher={Elsevier}
}

@article{raftopoulou2004cell,
  title={Cell migration: Rho GTPases lead the way},
  author={Raftopoulou, Myrto and Hall, Alan},
  journal={Developmental biology},
  volume={265},
  number={1},
  pages={23--32},
  year={2004},
  publisher={Elsevier}
}

@article{zegers2014rho,
  title={Rho GTPases in collective cell migration},
  author={Zegers, Mirjam M and Friedl, Peter},
  journal={Small GTPases},
  volume={5},
  number={3},
  pages={e983869},
  year={2014},
  publisher={Taylor \& Francis}
}

@article{levine2006directional,
  title={Directional sensing in eukaryotic chemotaxis: a balanced inactivation model},
  author={Levine, Herbert and Kessler, David A and Rappel, Wouter-Jan},
  journal={Proceedings of the National Academy of Sciences},
  volume={103},
  number={26},
  pages={9761--9766},
  year={2006},
  publisher={National Acad Sciences}
}

@article{duczek2019mass,
  title={Mass lumping techniques in the spectral element method: On the equivalence of the row-sum, nodal quadrature, and diagonal scaling methods},
  author={Duczek, S and Gravenkamp, H395973607189905},
  journal={Computer Methods in Applied Mechanics and Engineering},
  volume={353},
  pages={516--569},
  year={2019},
  publisher={Elsevier}
}

@article{gravenkamp2020mass,
  title={On mass lumping and explicit dynamics in the scaled boundary finite element method},
  author={Gravenkamp, Hauke and Song, Chongmin and Zhang, Junqi},
  journal={Computer Methods in Applied Mechanics and Engineering},
  volume={370},
  pages={113274},
  year={2020},
  publisher={Elsevier}
}

@article{wise2008three,
  title={Three-dimensional multispecies nonlinear tumor growth—I: model and numerical method},
  author={Wise, Steven M and Lowengrub, John S and Frieboes, Hermann B and Cristini, Vittorio},
  journal={Journal of theoretical biology},
  volume={253},
  number={3},
  pages={524--543},
  year={2008},
  publisher={Elsevier}
}

@article{bosgraaf2009ordered,
  title={The ordered extension of pseudopodia by amoeboid cells in the absence of external cues},
  author={Bosgraaf, Leonard and Van Haastert, Peter JM},
  journal={PloS one},
  volume={4},
  number={4},
  pages={e5253},
  year={2009},
  publisher={Public Library of Science San Francisco, USA}
}

@article{fuller2010external,
  title={External and internal constraints on eukaryotic chemotaxis},
  author={Fuller, Danny and Chen, Wen and Adler, Micha and Groisman, Alex and Levine, Herbert and Rappel, Wouter-Jan and Loomis, William F},
  journal={Proceedings of the National Academy of Sciences},
  volume={107},
  number={21},
  pages={9656--9659},
  year={2010},
  publisher={National Acad Sciences}
}

@article{hughes2005isogeometric,
  title={Isogeometric analysis: CAD, finite elements, NURBS, exact geometry and mesh refinement},
  author={Hughes, Thomas JR and Cottrell, John A and Bazilevs, Yuri},
  journal={Computer methods in applied mechanics and engineering},
  volume={194},
  number={39-41},
  pages={4135--4195},
  year={2005},
  publisher={Elsevier}
}

@article{dalcin2016petiga,
  title={PetIGA: A framework for high-performance isogeometric analysis},
  author={Dalcin, Lisandro and Collier, Nathaniel and Vignal, Philippe and C{\^o}rtes, AMA and Calo, Victor M},
  journal={Computer Methods in Applied Mechanics and Engineering},
  volume={308},
  pages={151--181},
  year={2016},
  publisher={Elsevier}
}

@book{emmerich2003diffuse,
  title={The diffuse interface approach in materials science: thermodynamic concepts and applications of phase-field models},
  author={Emmerich, Heike},
  volume={73},
  year={2003},
  publisher={Springer Science \& Business Media}
}

@article{szabo_mechanisms_2018,
    title = {Mechanisms of {Neural} {Crest} {Migration}},
    volume = {52},
    issn = {0066-4197, 1545-2948},
    url = {https://www.annualreviews.org/doi/10.1146/annurev-genet-120417-031559},
    doi = {10.1146/annurev-genet-120417-031559},
    language = {en},
    number = {1},
    urldate = {2025-01-03},
    journal = {Annual Review of Genetics},
    author = {Szabó, András and Mayor, Roberto},
    month = nov,
    year = {2018},
    pages = {43--63},
}
\biboptions{sort&compress}

\end{document}